\documentclass[sts]{imsart}

\RequirePackage{amsthm,amsmath,amsfonts,amssymb}
\RequirePackage[numbers]{natbib}
\usepackage{float}
\usepackage{pgf,tikz}
\usepackage{booktabs}
\usetikzlibrary{positioning}
\usetikzlibrary{arrows.meta}

\startlocaldefs
\endlocaldefs

\begin{document}
\usetikzlibrary{arrows}
\newcommand{\X}{\boldsymbol{X}}
\newcommand{\A}{\boldsymbol{A}}
\newcommand{\U}{\boldsymbol{U}}
\newcommand{\Y}{\boldsymbol{Y}}
\newcommand{\Z}{\boldsymbol{Z}}
\newcommand{\x}{\boldsymbol{x}}
\newcommand{\W}{\boldsymbol{W}}
\newcommand{\tA}{\Tilde{\A}}
\newcommand{\tX}{\Tilde{\X}}
\newcommand{\tY}{\Tilde{\Y}}
\newcommand{\ta}{\Tilde{a}}
\newcommand{\tx}{\Tilde{\x}}
\newcommand{\R}{\boldsymbol{\mathcal{R}}}
\newcommand{\ess}{\boldsymbol{S}}
\newcommand{\miota}{\boldsymbol{M}_{\boldsymbol{1}}}
\newcommand{\piota}{\boldsymbol{P}_{\boldsymbol{1}}}
\newcommand{\mz}{\boldsymbol{M_z}}

\newcommand{\one}{\boldsymbol{1}}
\newcommand{\bhatx}{\hat{\beta}_a}
\newcommand{\naive}{\hat{\beta}_a(\emptyset)}
\newcommand{\siga}{\sigma_a^2}
\newcommand{\sigu}{\sigma_u^2}
\newcommand{\sigx}{\boldsymbol{\sigma}_{\x}^2}
\newcommand{\sigy}{\sigma_y^2}
\newcommand{\gu}{\gamma_u}
\newcommand{\gx}{\boldsymbol{\gamma}_{\x}}
\newcommand{\bx}{\boldsymbol{\beta}_{\x}}
\newcommand{\ba}{\beta_{a}}
\newcommand{\bu}{\beta_{u}}
\newcommand{\eone}{\boldsymbol{\epsilon_{1}}}
\newcommand{\etwo}{\boldsymbol{\epsilon_{2}}}
\newcommand{\ay}{\alpha_{y}}
\newcommand{\au}{\alpha_{u}}
\newcommand{\ax}{\alpha_{x}}
\newcommand{\ama}{\boldsymbol{A^TM_zA}}
\newcommand{\amy}{\boldsymbol{A^TM_zY}}
\newcommand{\amu}{\boldsymbol{A^TM_zU}}
\newcommand{\amb}{\boldsymbol{A^TM_zBAV}}
\newcommand{\pto}{\overset{p}{\to}}
\newcommand{\mx}{\boldsymbol{M_X}}
\newcommand{\px}{\boldsymbol{P_X}}
\newcommand{\mxone}{\boldsymbol{M_{X_1}}}
\newcommand{\mxtwo}{\boldsymbol{M_{X_2}}}
\newcommand{\pxone}{\boldsymbol{P_{X_1}}}
\newcommand{\mxcj}{\boldsymbol{M_{X^c_j}}}
\newcommand{\sigaprime}{\sigma_a^{2\prime}}
 
\newcommand{\E}{\ensuremath{\mathrm{E}}}
\newcommand{\cov}{\mbox{cov}}
\newcommand{\cor}{\mbox{corr}}
\newcommand{\var}{\mbox{var}}
\newcommand{\VaR}{\mbox{VaR}}
\newcommand{\TVaR}{\mbox{TVaR}}
\newcommand{\fO}{f^{\mathcal{O}}}
\newcommand{\fE}{f^{\mathcal{E}}}
\newcommand{\EO}{E^{\mathcal{O}}}
\newcommand{\EE}{E^{\mathcal{E}}}
\newcommand{\covO}{Cov^{\mathcal{O}}}
\newcommand{\covE}{Cov^{\mathcal{E}}}

\newcommand{\varO}{Var^{\mathcal{O}}}
\newcommand{\varE}{Var^{\mathcal{E}}}
\renewcommand{\d}{\mbox{d}}
\def\qed{{\hfill $\Box$}\bigskip}

\newcommand{\pO}{\boldsymbol{P}^{\mathcal{O}}}
\newcommand{\pE}{\boldsymbol{P}^{\mathcal{E}}}

\newcommand{\pOn}{\boldsymbol{P}^{\mathcal{O}_n}}
\newcommand{\pEn}{\boldsymbol{P}^{\mathcal{E}_n}}
\newcommand{\C}{\boldsymbol{\mathcal{C}}}
\newcommand{\Q}{\boldsymbol{Q}}

\newcommand{\EOn}{E^{\mathcal{O}_n}}
\newcommand{\EEn}{E^{\mathcal{E}_n}}

 \newcommand{\zero}{\boldsymbol{0}}
\newcommand{\bbeta}{\boldsymbol{\beta}}
\newcommand{\btheta}{\boldsymbol{\theta}}
\newcommand{\bpsi}{\boldsymbol{\psi}}
\newcommand{\bgamma}{\boldsymbol{\gamma}}
\newcommand{\balpha}{\boldsymbol{\alpha}}
\newcommand{\sigmoid}{s}
\newcommand{\radial}{r}
\newcommand{\s}{\boldsymbol{s}}
\newcommand{\onea}{1\{\A =a\}}
\newcommand{\oneap}{1\{\A =a^\prime\}}
\newcommand{\argmax}{arg\,max}
\newcommand{\plim}{plim\,}
\newcommand{\plimn}{\underset{n\to\infty}{\plim}}
\newcommand{\sgn}{\text{sgn}}

\newcommand{\onep}{\underset{1\times p}{\one}}
\definecolor{pal1}{HTML}{003f5c}
\definecolor{pal2}{HTML}{7a5195}
\definecolor{pal3}{HTML}{ef5675}
\definecolor{pal4}{HTML}{ffa600}

\definecolor{virid1}{HTML}{ffa600}
\definecolor{virid2}{HTML}{35b779}
\definecolor{virid3}{HTML}{31688e}
\definecolor{virid_purp}{HTML}{440154}

\definecolor{darj_red}{HTML}{ff0000}
\definecolor{darj_teal}{HTML}{00A08A}
\definecolor{darj_yel}{HTML}{F2AD00}
\definecolor{darj_orange}{HTML}{F98400}
\definecolor{darj_blue}{HTML}{5BBCD6}

\definecolor{br_green}{HTML}{354823}
\definecolor{br_blue}{HTML}{273046}
\begin{frontmatter}
\title{Simulation Experiments as a Causal Problem}
%\title{A sample article title with some additional note\thanksref{T1}}
\runtitle{Simulation Experiments as a Causal Problem}
%\thankstext{T1}{A sample of additional note to the title.}

\begin{aug}
\author[A]{\fnms{Tyrel}~\snm{Stokes}\ead[label=e1]{tyrel.stokes@mail.mcgill.ca}},
\author[B]{\fnms{Ian}~\snm{Shrier}\ead[label=e2]{ian.shrier@mcgill.ca}}
\and
\author[C]{\fnms{Russell}~\snm{Steele}\ead[label=e3]{russell.steele@mcgill.ca}}
%%%%%%%%%%%%%%%%%%%%%%%%%%%%%%%%%%%%%%%%%%%%%%
%% Addresses                                %%
%%%%%%%%%%%%%%%%%%%%%%%%%%%%%%%%%%%%%%%%%%%%%%
\address[A]{NYU Langone, Department of Biostatistics\printead[presep={\ }]{e1}.}
\address[B]{McGill University, Department of Mathematics\printead[presep={\ }]{e2}.}
\address[C]{Lady Davis Institute\printead[presep={\ }]{e3}.}
\end{aug}

\begin{abstract}
Simulation methods are among the most ubiquitous methodological tools in statistical science. In particular, statisticians often is simulation to explore properties of statistical functionals in models for which developed statistical theory is insufficient or to assess finite sample properties of theoretical results. We show that the design of simulation experiments can be viewed from the perspective of causal intervention on a data generating mechanism. We then demonstrate the use of causal tools and frameworks in this context. Our perspective is agnostic to the particular domain of the simulation experiment which increases the potential impact of our proposed approach. In this paper, we consider two illustrative examples. First, we re-examine a predictive machine learning example from a popular textbook designed to assess the relationship between mean function complexity and the mean-squared error. Second, we discuss a traditional causal inference method problem, simulating the effect of 
unmeasured confounding on estimation, specifically to illustrate bias amplification. In both cases, applying causal principles and using graphical models with parameters and distributions as nodes in the spirit of influence diagrams can 1) make precise which estimand the simulation targets , 2) suggest modifications to better attain the simulation goals, and 3) provide scaffolding to discuss performance criteria for a particular simulation design. 

\end{abstract}


\end{frontmatter}
%%%%%%%%%%%%%%%%%%%%%%%%%%%%%%%%%%%%%%%%%%%%%%
%%%% Main text entry area:

\section{Introduction}

Simulation methods are among the most ubiquitous tools in statistical science. In particular, these methods are commonly used for checking and exploring properties of statistical functionals in contexts and models where developed statistical theory is insufficient. Even when proposing new theoretical results, researchers often use simulation to support or explore the theory, because in these contexts we can evaluate the theory with knowledge of the ground truth.  Similarly, asymptotic theoretical results may not be applicable without extremely large sample sizes or in high-dimensional models. Such simulations play a key role in both methodological development\citep{morris2019using} and applied modelling settings \citep{gelman2020bayesian}.\\

However, while crucial to modern approaches to statistics, statisticians do not always design simulations explicitly as they would a physical experiment. In this paper, we demonstrate that simulation experiments can be characterized as causal interventions on data generating mechanisms. From this perspective, one can leverage the tools and ideas of the field of  causal inference, such as graphical models and potential outcomes, to design more efficient, interpretable, and informative simulation experiments.\\

Of course, there already exists a large body of research on the design of computer experiments \citep{GARUD201771,sacks1989}.  Modern research in the design of computer experiments focuses primarily on utilizing realistic or high-fidelity simulators of physical systems, for example large meteorological or cosmological systems, which can generate synthetic data similar to that of their physical counterparts. The goal in that context is often to design optimal strategies to sample points along the input domain to generate maximally informative observation samples \citep{GARUD201771}. In particular, space-filling designs are created via various static and adaptive methods which have been proposed to leverage various features and external information about the system. The choices of which variables to intervene upon are considered fixed by the sampling domain, and the goal is the intelligently sample over this pre-determined sampling domain.\\

The perspective advanced in this paper is more analogous to the classical design of physical experiments, where choices are made to determine the nature and kind of variation observed due to different data generating mechanisms. That is, the problem is to define the sampling domain and over which variables and parameters we will induce variation and how. In this paper, we emphasize the importance of defining the estimand for simulation experiments and carefully interpreting which causal pathways the outcome variation represents (e.g direct and indirect effects). Our proposed approach remains agnostic to the underlying purpose of the simulation experiment and we show applications both in testing causal inference theory and in a predictive machine learning context. Once the estimand is well-defined, many statistical simulation experiments may be best suited to space-filling designs, particularly in complicated or high-dimensional domains or when it is desirable that the effect of interest be integrated over some choices of ancillary parameters, and statisticians ought to take advantage of this literature when appropriate.\\

Designing interpretable simulation experiments requires the ability to control for additional outside factors to create fair comparisons within
a structure for well-calibrated decision-making. Further, we can use the formal framework and tools of causal inference to help implement these adjustments and ensure that our statistical experiments are able to estimate the intended effect or effects of interest. For example, researchers already make explicit adjustments to the error terms in
their statistical experiments so that the outcomes may be more fairly compared but without the formal causal inference framework we propose here. We show that causal inference theory and tools can take these intuitions further and help better design these types of experiments and give additional layers of interpretability and understanding by framing their estimands in terms of causal effects,
as well as protect them from potential biases that result from seemingly sensible, but ultimately harmful modifications to the design. These ideas in this article are domain agnostic, in that they do not just apply to statistical simulation experiments. \\

\subsection{Elements of Statistical Learning Example}\label{esl sec}

In this section, we discuss an experiment provided by Hastie et al \cite{hastie2009elements}, which illustrates how varying the complexity of the mean function impacts the ability of a neural network to produce accurate predictions. Specifically, the authors compare the results from using a sum of sigmoids as the true mean function  to the results when using a product of radial basis functions as the truth. In the original experiment, they also varied the number of layers used in the fitted neural network, but for clarity we will keep the number of layers used fixed at 2. The two data generating processes for the outcome are described in equations \eqref{Y sigmoid} and \eqref{Y radial}:
\begin{align}
    \Y_{\sigmoid} &= \sum_{i=1}^{p_1}\sigma(\balpha_i^T\X) + \epsilon_{\sigmoid}\label{Y sigmoid}\\
    \Y_{\radial} &= \prod_{i=1}^{p_2}\phi(\X_i) + \epsilon_{\radial},\label{Y radial}
\end{align}
where $p_1$ is the number of standard normal variables $\X$ in the sigmoid experiment and $p_2$ is the number of standard normal variables in the radial experiment. In the original experiment, they fixed $p_1 =2$ and $p_2 = 10$. The function $\sigma(\cdot)$ is the sigmoid function and $\phi(\cdot)$ is the standard normal density function. In the original experiment $\balpha_1 = (3,3)$ and $\balpha_2 = (3,-3)$.\\

Hastie et al \citep{hastie2009elements} had the insight that a direct comparison between these two mean functions in terms of the mean-squared error (MSE) was unlikely to yield fair and comparable results. Without careful consideration, one might incorrectly attribute the results of the experiment to other important factors. In particular, the authors noted that the signal-to-noise ratios of the two data generating processes were not guaranteed to be equal or even comparable which could impact the recovered MSE. The authors opted to make two adjustments to the simulation experiment to control for this potential imablance across the two treatment arms.  First, they used mean-squared error relative to the Bayes Risk instead of MSE as the criterion for comparison and then they fixed the signal-to-noise ratio in both treatment arms by varying the noise to account for differences in the variance induced by the mean function. A priori, these adjustments make intuitive sense and are an enlightening example of leveraging authors' domain knowledge in machine learning to create more meaningful simulation experiments. By leveraging formal causal inference tools and frameworks in addition to the causal thinking demonstrated by the authors, we show here that we 1) precisely characterize the estimand the simulation targets, 2) suggest further modifications to better attain the simulation goals, and 3) provide scaffolding to discuss other
performance criteria for a particular simulation design, such as the generalizability to other
contexts.\\

In order to to frame the statistical experiment in terms of causal inference theory, we must 
first identify the outcome and treatment. As stated above, the outcome the authors use is the MSE relative to the Bayes Risk. In this additive linear setting, the Bayes Risk is equivalent to the irreducible error variance ($Var(\Y - E[\Y|\X])$), i.e $Var(\epsilon_{\sigmoid}) = \sigma^2_{\sigmoid}$ and $Var(\epsilon_{\radial}) = \sigma^2_{\radial}$ respectively. Although we will primarily be analyzing this statistical experiment from the perspective of potential outcomes, graphical models are useful and powerful tools for understanding simulation experiments and generating sets of control variables. We want to build a graph in terms of nodes which are either parameters which we can directly control (later called directly manipulable parameters $\theta_m$) or other important distributions or functionals which form confounding or mediating pathways. Much like standard causal inference, what we might consider a mediating pathway will depend upon the question we hope the simulation to answer. In this case, we are specifically trying to isolate the effect of the mean-function complexity and this will drive our approach to decomposing the outcome.\\

We now expand upon the standard three-way decomposition of the mean squared error used in Chapter 11 of The Elements of Statistical Learning \citep{hastie2009elements} in order to isolate the role of the signal-to-noise ratio in relative mean-squared error to understand the impact of holding it constant. The three-way decomposition breaks the mean-squared error into the irreducible error ($Var(Y - E[\Y|\X])$), the misspecification error ($E[(\mu(\X) - f(\X;\btheta_0))^2]$, where $\mu(\X) = E[\Y|\X])$ is the mean function of the data generating mechanism), and the model variance ($E[(\hat{f}(\X;\hat{\btheta}) - f(\X;\btheta_0))^2]$). We will call the optimal parameter $\btheta_0$ following \cite{buja2019models2}. The optimal parameter or parameters are those which minimize the loss function in the large data limit. When the loss function is the cross-entropy the optimal parameter set forms an equivalence class with respect to the implied density $f(\X;\btheta_0)$ \citep{watanabe2018mathematical}.  Although suppressed in the notation, the optimal parameter(s) typically depend on the distribution of the regressors unless the model is well-specified \cite{buja2019models2}. We will assume that the two layer neural network with a fixed number of parameters is not necessarily well-specified for both data generating processes in this case.\\

The misspecification error can be decomposed further in terms of the model signal, $Var(E[\Y|\X])$, as follows \cite{hastie2009elements}
\begin{align*}
    &E[(\mu(\X) - f(\X;\btheta_0)^2] = Var(E[\Y|\X]) + Var(f(\X;\btheta_0))\\
    &+ (E[f(\X;\btheta_0)] - E[Y])^2 - 2Cov(\mu(\X),f(\X;\btheta_0)).
\end{align*}

This expression allows us to express the mean square error relative to the Bayes risk (or irreducible error) directly in terms of the signal-to-noise ratio as follows:
\begin{align*}
    &\frac{E[(Y - \hat{f}(\X;\hat{\btheta}))^2]}{Var(\Y - E[\Y|\X])} = 1 + \frac{Var(E[\Y|\X])}{Var(\Y - E[\Y|\X])}+\\
    & \frac{E[(\mu(\X) - f(\X;\btheta_0)^2]}{Var(\Y - E[\Y|\X])} + \frac{Var(f(\X;\btheta_0))}{Var(\Y - E[\Y|\X])} +\\
    &\frac{(E[f(\X;\btheta_0)] - E[Y])^2 - 2Cov(\mu(\X),f(\X;\btheta_0))}{Var(\Y - E[\Y|\X])}+\\
    &
    \frac{E[(\hat{f}(\X;\hat{\btheta}) - f(\X;\btheta_0))^2]}{Var(\Y - E[\Y|\X])}.
\end{align*}

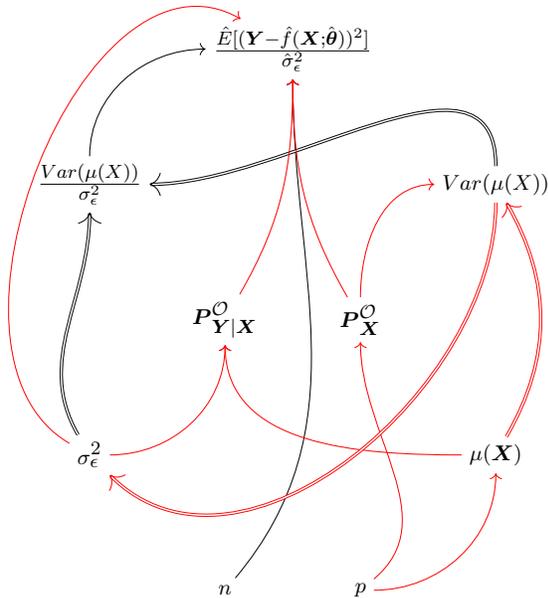
\begin{figure}[H]
\centering
\scalebox{.9}{
\begin{tikzpicture}
\node (n) at (-1,0) {$n$};
\node (p) at (1,0) {$p$};
\node (sig_ey) at (-3,2) {$\sigma_{\epsilon}^2$};
\node (mu_x) at (3,2) {$\mu(\X)$};
\node (px) at (1,4) {$\pO_{\X}$};
\node (p_yx) at (-1,4) {$\pO_{\Y|\X}$};
\node (varmux) at (3,6) {$Var(\mu(X))$};
\node (snr) at (-3,6) {$\frac{Var(\mu(X))}{\sigma_{\epsilon}^2}$};
\node (repe) at (0,8) {$\frac{\hat{E}[(\Y - \hat{f}(\X;\hat{\btheta}))^2]}{\hat{\sigma}_{\epsilon}^2}$};
%\draw[->] (n) to [out = 20, in = -90] (px);
%\draw[->] (n) to [out = 140, in = -90] (p_yx);
\draw[->] (n) to [out = 50, in = -90] (repe);
\draw[->,darj_red] (mu_x) to [out = -180, in =-90] (p_yx);
\draw[double,darj_red, ->] (mu_x) to [out = 60, in = -60] (varmux);
\draw[->,darj_red] (p) to [out = 40, in = -90] (px);
\draw[->,darj_red] (p) to [out = 0, in = -90] (mu_x);
\draw[double,->] (varmux) to [out = 90, in = 0] (snr);
\draw[->, double,darj_red] (varmux) to [out = 270, in = -45](sig_ey); 
\draw[double, ->] (sig_ey) to [out = 120, in = -90] (snr);
\draw[->,darj_red] (sig_ey) to [out = 0, in = -90] (p_yx);
\draw[->,darj_red] (sig_ey) to [out = -210, in = -210] (repe);
\draw[->,darj_red] (px) to [out = 90, in = -180] (varmux);
\draw[->,darj_red] (px) to [out = 120, in = -90] (repe);
\draw[->,darj_red] (p_yx) to [out = 60, in = -90] (repe);
\draw[->] (snr) to [out = 90, in = -180] (repe);
\end{tikzpicture}
}
\caption[Graph of ESL simulation experiment]{A graph of the experiment in Elements of Statistical Learning. Several of the edges in this graph are deterministic, and are shown with double arrows following conventions in \cite{arnold2020causal}. In this case we let $n = (n_{train},n_{test})$ for simplicity, but in principle they could be considered different nodes. In Section \ref{sc:sim exp} we will further discuss the principles under which this DAG was constructed. In brief, the parent nodes of the DAG are constructed with directly manipulable parameters which we control in the simulation data generation design. In this context we have the sample size ($n$) the number covariates ($p$) and the irreducible outcome error ($\sigma_{\epsilon}^2$) as the outermost manipulable parameters. The next layer of child nodes are more complicated functionals and distributions which we directly control and simulation, such as the conditional outcome mean $\mu(\X)$, the regressor distribution ($\pO_{\X}$), and subsequently the conditional outcome distribution $\pO_{\Y|\X}$. The paths which are highlighted in red are confounding pathways. The order of the DAG is explicitly related to the joint distribution decomposition we have chosen to simulate from as well as other experimental design choices. All other functionals further downstream on the DAG are functionals of the joint distribution $\pO_{\Y,\X}$. Sometimes we can use statistical theory to eliminate edges or to draw more precise pathways of interest as we have done in this case by decomposing the MSE-to-bayes-risk-ratio into an explicit function of the signal to noise ratio which is of interest in this particular simulation. In this experiment we vary the mean function and as such this is our treatment variable which the MSE-to-bayes-risk-ratio being the outcome.}
\label{fig: esl graph}
\end{figure}

In this particular case, the noise is exactly $\sigma_{\epsilon_i}^2$ where $i \in \{\sigmoid,\radial\}$ meaning either sigmoid or radial. This allows us to express the experiment outcome, sample relative mean-squared error, as $\hat{E}[(\Y - \hat{f}(\X;\hat{\btheta}))^2]$. We will assume that the sample relative mean-squared error depends on its mean (as described above) and the sample error, where the sample error depends only systematically on $n$. This allows us to draw a Directed Acyclic Graph (DAG) with a mediating pathway through the signal-to-noise ratio as seen in Figure \ref{fig: esl graph}. The nature of the nodes in the DAG is different than the traditional causal setting. However, we will explain below that the DAG belongs to the class of {\em influence diagrams} \cite{dawid2002influence}, which generalizes the notion of interventions beyond how DAGs have typically been used in causal inference.

The graph in Figure \ref{fig: esl graph} illustrates fully the consequences of holding the signal-to-noise ratio constant in the original experiment and whether it does, in fact, allow us to interpret the simulation results in terms of the pathway from mean-function complexity to the outcome. Figure \ref{fig: esl graph} displays several features complicating the original interpretation of the simulation as an experiment to assess the effect of data-generating distribution on the prediction performance of the neural networks.  First, there may be confounding pathways between the mean function complexity and the relative MSE in the original experiment design. Notice, for example, that there is a pathway from $p$, the number of covariates, to the treatment ($p \rightarrow \mu(\X))$ and also a pathway to the outcome through the regressor distribution which is not blocked by holding the signal-to-noise ratio constant ($p \rightarrow \pO_{\X}\rightarrow \frac{\hat{E}[(\Y - \hat{f}(\X;\hat{\btheta}))^2]}{\hat{\sigma}_{\epsilon}^2})$. Although all of the regressors ($\X$) in both treatments are marginally independent standard normal the joint regressor distribution ($\pO_{\X}$) changes as the number of covariates is varied. This modifies the misspecification error through the signal-to-noise ratio, but also independently through the optimal parameter $\btheta_0(\pO_{\X})$.\\

Most importantly, varying the number of covariates changes the magnitude of the signal and thus the irreducible error required to hold the SNR constant at 4, since in this experiment design the noise is a deterministic function of the signal. With high probability the  radial function takes values which are quite small (<0.1) and taking the product of 10 radials compared to, say, 2 can have a large impact on the order of the signal and thus the order of the irreducible error in this particular experiment design. This is problematic since the irreducible error is also on several confounding pathways as seen by the red paths in Figure \ref{fig: esl graph}. By changing the outcome from the MSE to the relative MSE this introduces additional pathways from the irreducible error to the outcome which are not through the signal-to-noise ratio. For example, the irreducible error now modifies the the model variance error and the remaining misspecification error outside of the signal-to-noise ratio as will be seen in equations \eqref{estimand} through \eqref{rem ms}.\\

In the original experiment by the ESL authors, the results were discussed as evidence of the effect of a difference in mean function complexity, but as we see in the graph the results were confounded. Of course, by modifying the definition of the treatment to include the number of covariates and the irreducible error the problem can be conceptually by-passed but it certainly makes the interpretation in terms of solely mean-function complexity significantly more difficult.\\

It should be noted that the graph in Figure \ref{fig: esl graph} features deterministic arrows and parameters as nodes, much like influence diagrams \citep{dawid2002influence}, to accommodate holding the population level property signal-to-noise ratio constant. To avoid the additional complications arising from applying standard d-separation rules in this context we will directly analyze the potential outcomes and derived estimand from the simulation experiment.\\

Let $\hat{MSE}_{rel} = \frac{\hat{E}[(\Y - \hat{f}(\X;\hat{\btheta}))^2]}{\hat{\sigma}_{\epsilon}^2}$ and $SNR = \frac{Var(\mu(\X))}{\sigma_{\epsilon}^2}$. The probability limit of the difference in relative mean squared error between the sigmoid and radial simulation experiments is:
{\scriptsize
\begin{align}
&\frac{\hat{E}[(\Y - \hat{f}(\X^{p_1};\hat{\btheta}_{\sigmoid}))^2]}{\hat{\sigma}_{\epsilon_{\sigmoid}}^2} - \frac{\hat{E}[(\Y - \hat{f}(\X^{p_2};\hat{\btheta}_{\radial}))^2]}{\hat{\sigma}_{\epsilon_{\radial}}^2} \pto\label{estimand}\\
   &E[MSE_{rel}|\mu(\X) = \mu_{\sigmoid}(\X^{p_1}),SNR =s,p_{1}, n,\sigma_{\epsilon_{\sigmoid}}]\nonumber\\
   &-  E[MSE_{rel}|\mu(\X) = \mu_{\radial}(\X^{p_2}),SNR =s,p_{2}, n,\sigma_{\epsilon_{\radial}}]\nonumber\\
   &= \left(\frac{E[(\mu_{\sigmoid}(\X^{p_1}) - f(\X;\btheta_0^{\sigmoid}))^2]}{\sigma_{\epsilon_{\sigmoid}}^2} -\frac{E[(\mu_{\radial}(\X^{p_2}) - f(\X;\btheta_0^{\radial}))^2]}{\sigma_{\epsilon_{\radial}}^2}\right)+\label{esl cond mean rec}\\
&\left(\frac{(E[f(\X;\btheta_0^{\sigmoid})] - E[\Y_{\sigmoid}])^2}{\sigma_{\epsilon_{\sigmoid}}^2} -\frac{(E[f(\X;\btheta_0^{\radial})] - E[\Y_{\radial}])^2}{\sigma_{\epsilon_{\radial}}^2} \right)+\label{esl mean rec}\\
   &\left(\frac{E[(\hat{f}(\X^{p_1};\hat{\btheta}_{\sigmoid}) - f(\X^{p_1};\btheta_0^{\sigmoid}))^2]}{\sigma_{\epsilon_{\sigmoid}}^2} - \frac{E[(\hat{f}(\X^{p_2};\hat{\btheta}_{\radial}) - f(\X^{p_2};\btheta_0^{\radial}))^2]}{\sigma_{\epsilon_{\radial}}^2}\right)\\
   &+ \left(\frac{Var(f(\X^{p_1};\btheta_0^{\sigmoid})) - 2Cov(\mu_{\sigmoid}(\X^{p_1}),f(\X^{p_1};\btheta_0^{\sigmoid}))}{\sigma_{\epsilon_{\sigmoid}}^2}\right.\nonumber\\
   &- \left.\frac{Var(f(\X^{p_2};\btheta_0^{\radial})) - 2Cov(\mu_{\radial}(\X^{p_2}),f(\X^{p_2};\btheta_0^{\radial}))}{\sigma_{\epsilon_{\radial}}^2}\right)\label{rem ms}.
\end{align}
}%
The above causal contrast is not easily interpretable at least in terms of the mean function complexity as we see four terms (lines \eqref{esl cond mean rec}-\eqref{rem ms}) which are modified by the varying irreducible error and, in some cases, also the regressor distribution. In the first line of equation \eqref{esl cond mean rec} we have a comparison of how well the neural network recovers the respective conditional means, but divided by the irreducible errors. Similarly, the next line, equation \eqref{esl mean rec}, tells us something about how well the models can recover the true mean but also modified by the standard errors.\\

In the original experiment they chose the irreducible error variance such that the signal-to-noise ratio was held constant at 4. In our re-creation of this experiment, we found that this requires $\sigma_{\epsilon_{\sigmoid}} \approx 2.9\times 10^{-1}$ and $\sigma_{\epsilon_{\radial}} \approx 2.9\times 10^{-6}$ which differ by many orders of magnitude. This means that all of the remaining terms for the model trained on radial data are increased significantly since they
 divided by the very small radial irreducible error. Again, this calls into question whether the simulation experiment tells us something important about the complexity of the mean functions and neural networks ability to recover the structure or whether the results are driven by differences in covariances and error terms.

\begin{figure}[H]
    \centering
    \includegraphics[scale = 0.3]{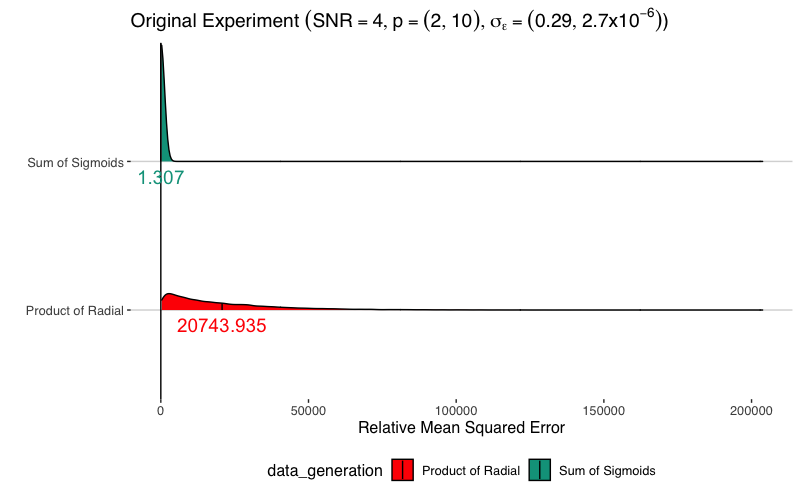}
    \caption[ESL simulation recreation]{Original experiment re-creation comparing the sum of sigmoid and product of radial mean functions. The sum of sigmoids has two regressors, whereas the product of radials has 10. The irreducible error variance was chosen so that the signal-to-noise ratio is held constant at 4. The models were fit on 100 training points and the relative mean-squared error was calculated on 10000 test training points as per the original test. The number of nodes was fixed to 2. The decay rate was set to 0.0005. The simulation experiment was repeated 5000 times, each time drawing a new sample. The original experiment only used a single training set.}
    \label{fig:esl orginal}
\end{figure}

\begin{table}[H]
\centering
\caption{Simulation Parameters for experiment in Figure \ref{fig:esl orginal}}
\begin{tabular}{c|ccc}
                       & \multicolumn{3}{c}{\textbf{Parameter}}                                              \\ \hline
\textbf{Mean Function} & $\sigma_{\epsilon}$                                   & $Var(E[\Y|\X])$       & SNR \\ \hline
Sum of Sigmoids        & 0.2858                                                & 0.327                 & 4   \\
Product of Radials     & $2.86\times 10^{-6}$ & $3.27\times 10^{-11}$ & 4  
\end{tabular}
\end{table}
In Figure \ref{fig:esl orginal} we see that the product of radials performs much worse according to the outcome by many orders of magnitude. This is a much more dramatic result than observed in the original publication even though we used the same methods. We suspect there might have been a typographical error for the number of covariates in the radial case in the textbook, but we were unable to get clarification from the authors. When the number of covariates is 2, the results are much closer to those presented in the textbook (see Figure \ref{fig: p2 =2} in Appendix \ref{app:esl fig}). It seems likely that much of this result is driven by the drastically different irreducible errors required to keep the signal-to-noise ratio constant and not simply a reflection of the complexity of fitting the product of radials mean function with a neural network.\\

It is important to note that this re-examination of the original simulation experiment doesn't simply muddy the interpretation, but also clarifies what kinds of adjustment might be necessary to accomplish the original simulation goals. A simple step to fix the confounding due to the covariate distribution would be to fix the number of covariates in both treatments to be two. We can also hold the signal-to-noise ratio fixed, but instead of varying the irreducible error which we know to have other important pathways to the outcome we can modify the parametrization of the sigmoid mean-function such that we fix the signal constant in both treatments. This requires changing $\balpha$ in the sigmoid condition from $(\balpha_1 = (3,3), \balpha_2 = (-3,3)$ to $(\balpha_1 \approx (.1,.1), \balpha_2 \approx (-.1,.1)$. With the signal fixed, we only need to choose the noise to be equal in both treatments for the signal-to-noise ratio to be constant. If we would like the signal-to-noise ratio to equal 4 as in the original experiment, we need to set the irreducible error to approximately 0.023. The results of this modified experiment are shown in Figure \ref{fig:2 cov,varmux}.

\begin{figure}[H]
    \centering
    \includegraphics[scale=0.3]{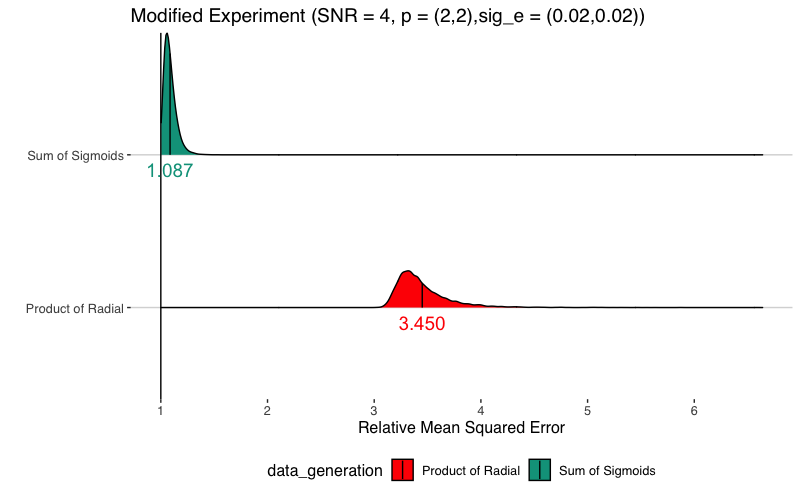}
    \caption[Modified ESL simulation holding regressor distribution, signal, and SNR constant]{Modified simulation experiment, where $p_2 =2$ and the parameters in the sum of sigmoids exposure are changed so that the variance of the conditional mean is constant in both arms of the treatment.}
    \label{fig:2 cov,varmux}
\end{figure}

In Figure \ref{fig:2 cov,varmux} we see that the product of radials still performs much worse than the sum of sigmoids, but this difference is now interpretable with respect to how well the neural network is able to capture the mean-function. There are of course caveats here as well as there are in most causal contexts. The estimand of this experiment is a conditional average treatment effect (CATE), where the conditioning is done at the values we have fixed. That is, we have evidence that the radial mean function results in the neural network not recovering the relative mean-squared error as well, but we have only shown this conditional on a single, particular parametrization. This framing however, offers further suggestions as to what might be required to collect evidence of the more general claim about the mean-function complexity. One approach might be to imagine a meaningful distribution of parameters and to average over these parameters to arrive at an unconditional Average Treatment Effect (ATE). In practice, this may be difficult to arrive at such a distribution of parameters that relevant communities would agree to. From this perspective we might also think of this problem in terms of generalizability and transportability \cite{bareinboim2013general,bareinboim2016causal} which gives a formal way to think about when and how results from one context may be transported or used in another context. Some of the skepticism about simulation studies might stem from disagreements about how and where the study might generalize to and hopefully formalizing simulations as a causal problem can lead to more fruitful discussions in this regard.\\

Two alternatives to targeting a particular CATE as the simulation estimand are 1) to build publicly available software which allows users to specify their own distribution of parameters, allowing them to choose an estimand targeted to their particular needs and 
2) to estimate a more general function for the CATE. Here we show an example of what this latter approach might look like in a paper.\\

One parameter we might be particularly worried about is the role of the irreducible error in driving the results in Figure \ref{fig:2 cov,varmux}. To mitigate this we might design an experiment which holds the signal constant as we did before, but varies the irreducible noise over a reasonably large range, for example (0.01,0.3).  All other parameters are taken to be identical to those used to create Figure \ref{fig:2 cov,varmux} and in particular the variance of the conditional mean. Therefore the signal-to-noise ratio, remains equal in the two treatments at each level of $\sigma_{\epsilon}$.

\begin{figure}[H]
    \centering
    \includegraphics[scale = 0.3]{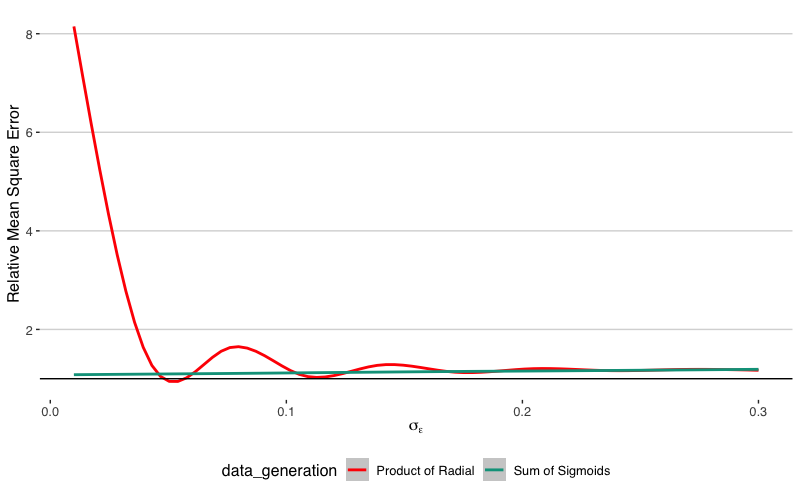}
    \caption[RMSE of radial and sigmoid data generating processes as irreducible noise varies] {This shows the relative mean squared error against the irreducible standard deviation. This experiment was done by discretizing (0.01,0.3) into a fine grid with size 50,000 and fitting a smoothed curve to the results.}
    \label{fig:mse by sigma}
\end{figure}
In Figure \ref{fig:mse by sigma} we see that for very small $\sigma_{\epsilon}$ the product of radials does much worse, but as the noise becomes large the two curves coalesce. Were we to take the signal to noise ratio to be 0.025 (i.e $\sigma_{\epsilon_{sigmoid}} = \sigma_{\epsilon_{radial}} = 0.2858$) we might conclude that there is no meaningful difference between the two mean functions, when globally this is not necessarily true. This particular experiment can be seen in Figure \ref{fig:snr 025} in Appendix \ref{app:esl fig}. It is important to remember again that these simulation results are conditional on the other parameters and only valid across $\sigma_{\epsilon} \in (0.01,0.3)$ since that is the experiment that was performed. Extrapolation beyond this range should only occur if we have strong theory which would allow us to generalize the effects.\\

\begin{figure}[H]
    \centering
    \includegraphics[scale = 0.3]{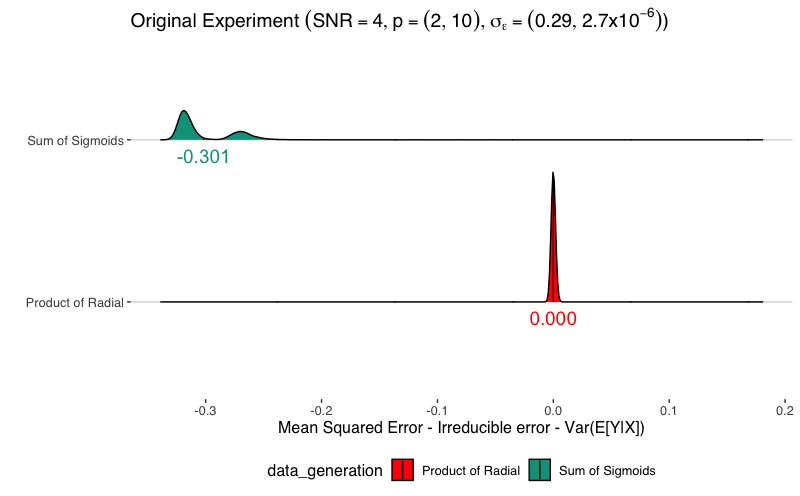}
    \caption{This is the original experiment with a modified outcome that no longer directly depends on the irreducible error and the signal.}
    \label{fig:esl original modout}
\end{figure}

In the previous batch of experiments we modified some of the parameters in the sum of sigmoid exposure in order to make it more comparable to the product of radials. In some cases, we might be specifically interested in the original parameterizations, notable $\balpha_1 = (3,3)$ and $\balpha_2 = (-3,3)$. In this case, we might have to find either a different way of conditioning on the parts of the outcome we are not interested in to conduct our experiment. One way to do that in this case would be to subtract off the irreducible error and the variance of the conditional mean of the mean squared error, instead of looking at the relative quantity. This leaves only the model variance error and the parts of the model misspecification error which do not depend on the variance of the conditional mean. Consider the original experiment with this new outcome in Figure \ref{fig:esl original modout}.
In Figure \ref{fig:esl original modout} we see that the sum of sigmoids performs much better than the product of radials. We know that this result does not depend on the scale of the irreducible error except through its effects on the conditional distribution and how the model is able to recover it. The fact that the sigmoid does so consistently well is likely related to the fact that it is well-specified by a two node neural network with a sigmoid basis function. This leaves only the model variance term in the model which varies. Since this model has few weights relative to the training data size, the model variance term should also be relatively small.\\

This example illustrates several points which we will expand on in the remaining article. First, adjustments to simulation experiments are sometimes necessary to fairly estimate the desired estimand. In some cases, the adjustments we make can have large effects on the outcome of the simulation study and lead to misleading or less meaningful conclusions. Second, the idea that one might make such modifications is in some cases quite natural. The authors of Elements of Statistical Learnings \cite{hastie2009elements}, Hastie et al, are experts in machine learning and understood that a naive comparison may be misleading. Third, it shows that formal tools from the existing causal inference literature, like graphical models and the potential outcomes framework can help us to make the goals of a simulation study precise and help guide the experimental design choices we make to estimate the desired estimand. In much the same way causal inference attempts to do this in the domains of medicine and epidemiology, the domain knowledge from statistical theory and experts in statistical methods can be leveraged to design better and more meaningful simulation experiments. Finally, casting simulation experiments as estimators for Conditional Average Treatment Effects (CATE) provides a clearer starting point to discuss the limitations of particular simulation studies while also suggesting routes to overcome those limitations in specific cases.

\section{Bias Amplification Intro}

To expand the theory of simulation experiments as a causal problem, we introduce the problem of bias amplification in linear models. We show that a causal perspective on simulation design in this context helps us to better interpret simulation experiments and resolve tensions between previous theoretical and simulation findings. In particular, theoretical analyses \citep{pearl2012class,pearl2011invited} noted that bias amplification may be potentially severe even when the bias amplifying variables are confounders, whereas the only large simulation study suggested that bias amplification is largely not an issue outside of the instrumental variable case \citep{meyers2011}. Below we outline the necessary background for the bias amplification problem, largely following \cite{stokes2022causal}. Consider the following linear structural equations (equations \eqref{Y truth} and \eqref{A truth}) which are assumed compatible with the DAG in Figure \ref{pearl DAG ch 5}.

\begin{figure}[H]
\centering
\begin{tikzpicture}
\node[text centered] (a) {$A$};
\node[right = 2.5 of a, text centered] (y) {$\boldsymbol{Y}$};
\node[above=1.5 of y, text centered] (x) {$\boldsymbol{X}$};
\node[below=1.5 of y, text centered] (u) {$\boldsymbol{U}$};
\draw[->, line width= 1] (a) --  (y);
\draw [->, line width= 1] (x) --  (y);
\draw [->, line width= 1] (u) --  (y);
\draw [->, line width= 1] (x) --  (a);
\draw [->, line width= 1] (u) --  (a);
\end{tikzpicture}
\caption{True data generating DAG}\label{pearl DAG ch 5}
\end{figure}
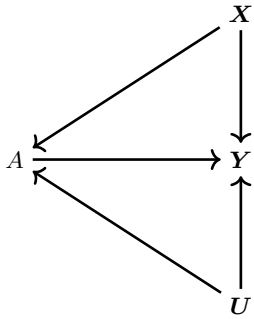
\begin{align}
\boldsymbol{Y} &= \alpha_y + \boldsymbol{A}\beta_a + \boldsymbol{U}\beta_u +\boldsymbol{X}\boldsymbol{\beta_{x}} + \boldsymbol{\epsilon_y}\label{Y truth}\\
\boldsymbol{A} &= \alpha_a + \boldsymbol{U}\gamma_u + \boldsymbol{X}\boldsymbol{\gamma_{x}} + \boldsymbol{\epsilon_a}\label{A truth}
\end{align}
Let $\Y$ be the outcome, $\A$ the treatment or exposure, and $(\X,\U)$ are confounding variables forming the only sufficient set on the DAG in Figure \ref{pearl DAG ch 5}. Throughout this section we will distinguish the set of variables $\X$ from $\U$ by assuming that the researcher has access to the vector $\X$, whereas $\U$ are a set of unmeasured variables. Since $\U$ is an unmeasured confounder, whenever $\gu,\bu \neq 0$ any feasible estimator, i.e an estimator which is a known functional of observed data, will be biased. The goal in this context is to understand how one might choose between feasible estimators to minimize bias.\\

Let $\bhatx(\ess)$ be the estimated OLS coefficient for the treatment, $\A$, on the outcome $\Y$ conditional on the set of variables $\ess$. For simplicity we assume throughout only linearly additive regression function specifications, i.e that $\bhatx(\ess)$ is the regression coefficient from the regression $\Y \sim \A + \ess$. Consider two sets of observable control variables $\ess_1,\ess_2$ on the same probability space $(\Omega,\mathcal{F},P)$ such that $\ess_1 \subset \ess_2$. In the linear structural equation context we say that bias amplification occurs when:
\begin{align}
|E[\bhatx(\ess_1)] - \beta_a| \leq |E[\bhatx(\ess_2)] - \beta_a|,
\end{align}
or asymptotically when:
\begin{align}
|\plimn\bhatx(\ess_1) - \beta_a| \leq |\plimn\bhatx(\ess_2) - \beta_a|.
\end{align}
We say that the additional variables in $\ess_2$, $\ess_2\setminus\ess_1$, are bias amplifying variables \citep{stokes2022causal} since by adding them to the conditioning set we increase the overall bias of the estimator. In this canonical example we are interested in comparing the estimator conditional on $\X$, $\bhatx(\X)$, to the naive estimator, $\bhatx(\emptyset)$. From \cite{stokes2022causal}, we have closed-form probability limits for the two estimators,
\begin{align}
\plimn (\bhatx(\X) - \beta_a) &= \frac{\bu\gu\sigu}{\siga - \gx^T\Sigma_{\X}\gx}\label{bhatx plim}\\
\plimn (\bhatx(\emptyset) - \beta_a) &= \frac{\bu\gu\sigu}{\siga} + \frac{\bx\gx\sigx}{\siga}.\label{naive plim}
\end{align}
The probability limits in equations \eqref{bhatx plim} and \eqref{naive plim}, in addition to the simplifying assumption that the $\X$'s are mutually independent,  give rise to the following expression characterizing the occurrence of bias amplification in this context:
\begin{align}
    &\frac{|\plimn (\bhatx(\X) - \beta_a)|}{|\plimn (\bhatx(\emptyset) - \beta_a)|} =\nonumber\\
    &\left( \frac{\siga}{\siga - \gx^T\sigx\gx}\right)\left(\frac{|\bu\gu\sigu|}{|\bu\gu\sigu + \onep(\bx\odot\gx\odot\sigx)|} \right) > 1,\label{eq bamp 1}
\end{align}
where $\odot$ is the Hadamard product or element-wise product. When $\bx,\gx,\sigx$ are $p$ dimensional vectors, or $p\times 1$ dimensional matrices, we have that $\bx\odot\gx\odot\sigx$ is a $p\times 1$ dimensional vector with typical element $(\bx\odot\gx\odot\sigx)_i = \beta_{x_i}\gamma_{x_i}\sigma_{x_i}^2$.\\

A natural simulation experiment in the bias amplification context is to better understand the effect of unmeasured confounding on bias amplification. In a sensitivity analysis, for example, we might ask how much unmeasured confounding is necessary for the bias of a particular estimator to cross some threshold (in the spirit of E-values \cite{VanderWeelePeng_Sensitivity2017} for example). Given the structural equations \eqref{Y truth} and \eqref{A truth}, the most straightforward way of translating the question "How does unmeasured confounding impact the bias of $\bhatx(\X)$ (or bias relative to $\bhatx(\emptyset)$)?" into a simulation design is to identify the structural equation parameters responsible for unmeasured confounding and then simulate over some distribution of those parameters.\\

In Figure \ref{pearl DAG ch 5} there are two causal pathways through which the unmeasured confounder contributes to bias, $\U \rightarrow \Y$ and $\U \rightarrow \A$ respectively. Suppose we are interested in isolating the effect of the unmeasured confounder through the treatment, $\U \rightarrow \A$. Inspecting the treatment structural equation \eqref{A truth} the parameter which controls the unmeasured confounder and treatment relationship is $\gu$. To assess the impact of changing $\U \rightarrow \Y$ one might run a simulation holding all parameters fixed and varying $\gu$ over some range of values. For simplicity we will discuss the experiment of changing $\gu$ from some reference value to $\gu^\prime$.\\

\begin{figure}[H]
    \centering
    \includegraphics[scale = 0.25]{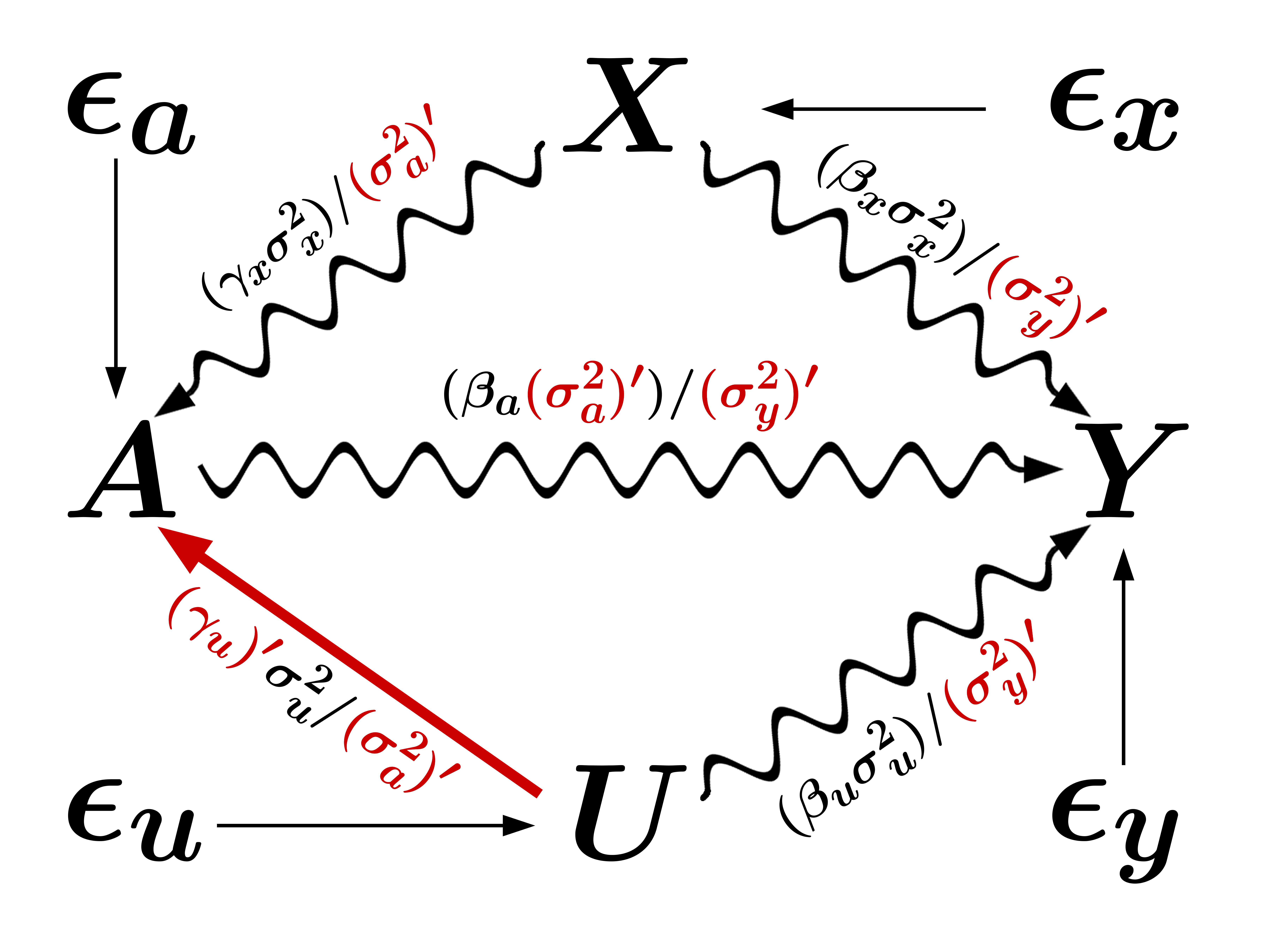}
    \caption{Extended DAG with error terms included visualizing the impact of changing $\gu$ to $\gu^\prime$. All edges where the proportion of variance explained has changed are squiggly except the intended edge of intervention ($\U \rightarrow \A$) which is shown in bold red. All quantities which have been changed are displayed in red. First appeared in \cite{stokes2022causal}.}
    \label{fig:ext dag 2}
\end{figure}

\begin{figure}[H]
    \centering
    \includegraphics[scale = 0.25]{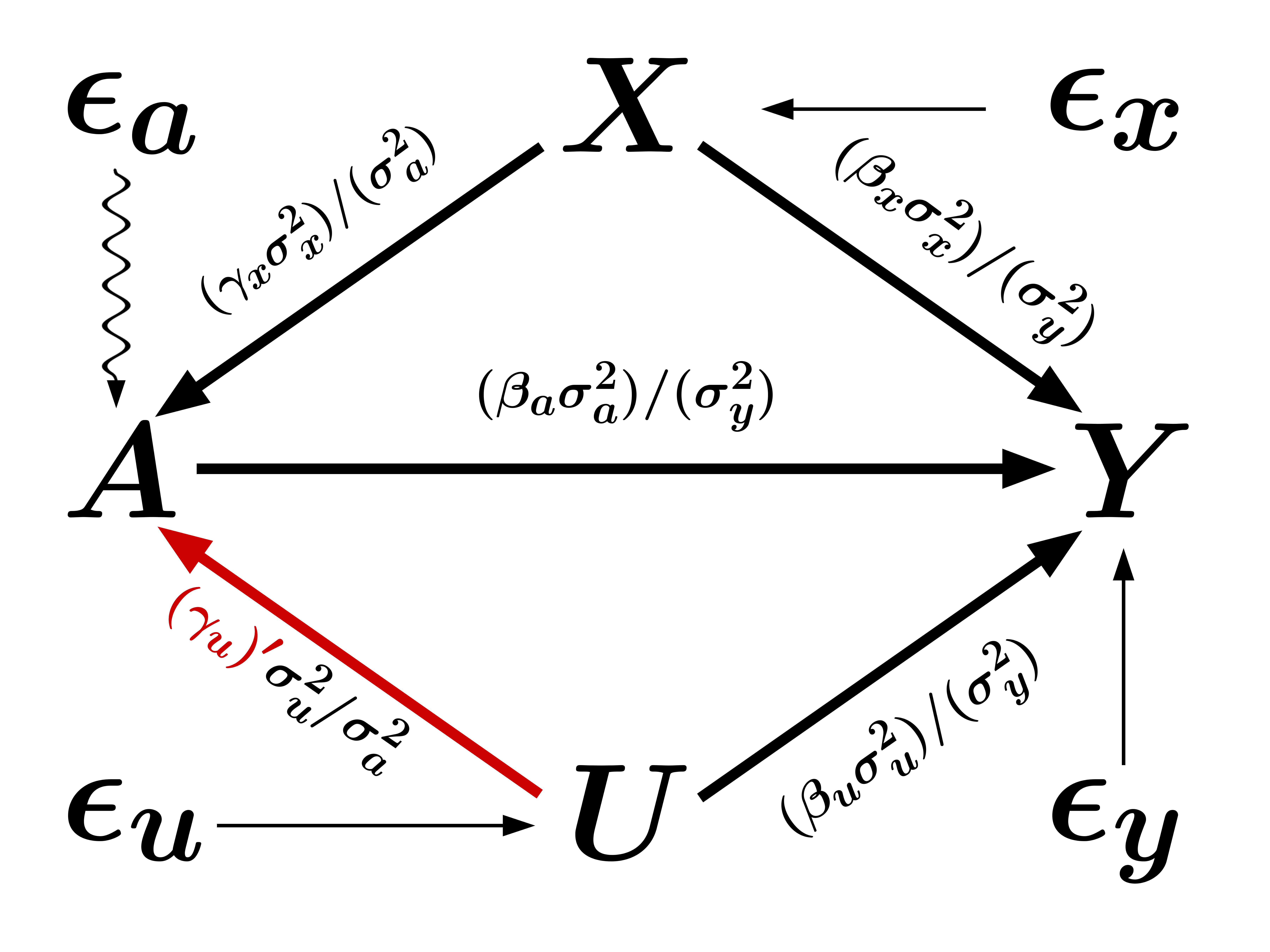}
    \caption{Extended DAG with error terms included visualizing the impact of changing $\gu$ to $\gu^\prime$ while using $\epsilon_a$ and $\epsilon_y$ to absorb changes in the marginal outcome and treatment variances. First appeared in \cite{stokes2022causal}.}
    \label{fig:ext dag 3}
\end{figure}

In \cite{stokes2022causal}, we argued that this kind of simulation can be misleading when assessing the impacts of bias amplification. Here we sketch the essentials of this argument before re-examining the simulation with the tools of causal inference. Broadly, the perspective taken in \cite{stokes2022causal} is that in linear models with additive effects and no interactions the proportion of variance explained parametrizes the strength of the edges in the graph. This is similar to the partial $\R^2$ characterization of omitted variable bias taken in \cite{cinelli2020making}. We can visualize this in the extended graph from \cite{stokes2022causal} reproduced below in Figure \ref{fig:ext dag 1}.
\begin{figure}[H]
    \centering
    \includegraphics[scale = 0.25]{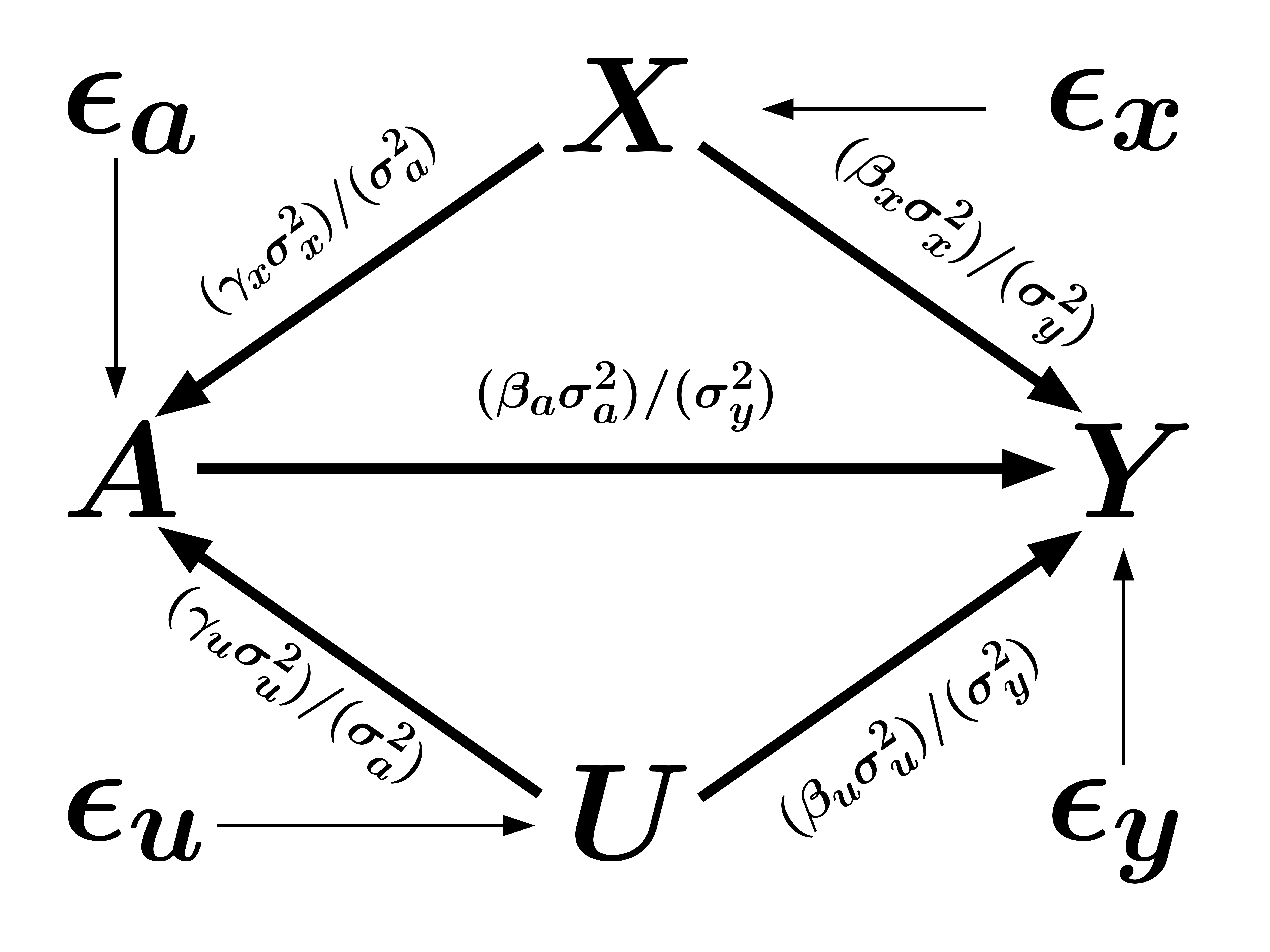}
    \caption{Extended DAG with error terms included. The formulas on each edge show the proportion of variance explained by the parent of the child node. First appeared in \cite{stokes2022causal}.}
    \label{fig:ext dag 1}
\end{figure}
When one simply varies the parameter $\gu$ in the structural equation, the proportion of treatment and outcome variance explained by the other variables, namely $\X$, is not held constant since this changes the marginal outcome and treatment variance. Most importantly in this case, the amount of treatment variance explained by $\X$ decreases as the magnitude of $|\gu|$ increases. Treatment variance explained by the covariates plays a crucial role in bias amplification, since it increases the asymptotic bias of the conditional estimator hyperbolically \citep{stokes2022causal,pearl2011invited}. This can be seen by examining equation \eqref{bhatx plim} and the first term in equation \eqref{eq bamp 1}. In other words, when we intervene directly on the parameter $\gu$, the total effects include (1) increasing the unmeasured confounding, and also (2) potentially decreasing the effect of the measured variables. The problem is not the simulation itself, but potentially the interpretation we assign the simulation results. We illustrate this in the extended DAG in Figure \ref{fig:ext dag 2}.
In Figure \ref{fig:ext dag 2} we see that the strength of the relationship between all non-error variables has been altered in terms of proportion of variance explained. This certainly results in a valid simulation experiment, but it is not clear that it answers the question "How does changing the unmeasured confounding through the treatment pathway change the bias of $\bhatx(\X)$ (or bias relative to $\bhatx(\emptyset)$)?" since we are intervening simultaneously on all of the edges. In \cite{stokes2022causal} we argued that we can remedy the simulation to answer the question at hand by using the variance of the error terms, $\epsilon_a$ and $\epsilon_y$, to absorb the changes to the marginal treatment and outcome variance induced by changing $\gu$. This allows us to hold the proportion of variance explained constant for all edges in the DAG except for $\U \rightarrow \A$ which we intended to intervene on. This modified experiment is visualized in the extended DAG in Figure \ref{fig:ext dag 3}.
Using the error term in this manner solves the simulation problem at hand and allows us to better understand bias amplification in this context at the cost of induced constraints on the parameter space. It does not, however, offer a clear path forward for how such simulations should be carried out in more complicated contexts where, say, given the models and structural equations the relationships between variables is not easily or appropriately expressed as proportion of variance. However, we argue that the underlying causal reasoning for making these simulations can be generalized to a broader context. The reason the original simulation depicted in Figure \ref{fig:ext dag 2} did not answer the question was that when varying the edge of interest we induced simultaneous variation in the data generating process which had an impact on our outcome (bias or bias amplification). This is a form of confounding in the experiment design. This problem was fixed by designing an intervention on the data generating mechanism which avoids the confounding variation. By re-framing a simulation experiment to answer a particular causal question related to the data generating mechanism, we can leverage existing causal theory to design better simulation experiments in contexts well beyond this specific example about bias amplification in linear models.

\section{Simulation Experiments as Causal Interventions}\label{sc:sim exp}

First we show how one might build a DAG for a simulation manipulating the strength of $\gu$. When constructing a graph for a simulation experiment, we first determine the set of parameters which we directly control or set in our experiment. These parameters need not be unique and will depend on the particular parametrizations that we choose, the functional form of any structural equations, as well as the joint distribution of the variables determined outside of the structural equations. These parameters will typically form the first set of parents in the graphical representation. In our canonical example, these quantities are all of the parameters in the two structural equations ($(\alpha_y,\beta_a,\beta_u,\bx)$ and $(\alpha_a,\gamma_u,\gx)$) and the means $(\mu_u,\boldsymbol{\mu}_{\x})$ and variances $(\sigu,\sigu,\sigx,\sigma_{\epsilon_y}^2,\sigma_{\epsilon_y}^2)$ of the variables we generate prior to the variables constructed through the structural equations. In this particular case all these pre-structural equation variables are generated with univariate normal distributions. The sample size $n$ is another manipulable parameter, which will combine with the other parameters to determine the empirical distribution for the realized random variables. Since these parameters are set by the user, they cannot be caused by other parameters or variables. In the simple case where these variables are set to constants they have degenerate distributions.\\

Given the set of manipulable parameters, the conditional distribution of the variables are determined by the structural equations. In this case, given the manipulable parameters, the conditional distributions for the outcome and the treatment are determined (equations \eqref{Y truth} and \eqref{A truth}). Then given these conditional distributions and the joint representation of the variables simulated outside of the structural equation, the joint distribution of all random variables are determined. We should think of outcomes or target variables for the simulation experiment as statistical functionals of this joint distribution. When we have available theory as in this case, we can take advantage of the more explicit representations of the functionals, perhaps in terms of the manipulable variables, to help us draw the simulation graph in more specific ways.\\

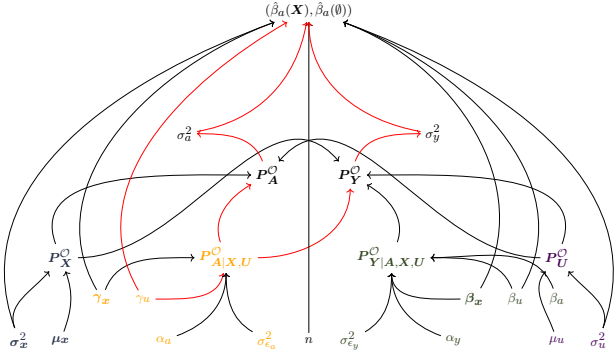
\begin{figure}[H]
\centering
\scalebox{0.55}{
\begin{tikzpicture}
\node[text = br_blue] (sig_x) at (-7,0) {$\sigx$};
\node[text = br_blue] (mu_x) at (-6,0) {$\boldsymbol{\mu_x}$};
\node[text = virid_purp] (sig_u) at (7,0) {$\sigu$};
\node[text = virid_purp] (mu_u) at (6,0) {$\mu_u$};
\node[text = br_green] (sig_ey) at (1,0) {$\sigma_{\epsilon_y}^2$};
\node[text = virid1] (sig_ea) at (-1,0) {$\sigma_{\epsilon_a}^2$};
\node[text = br_green] (bx) at (4,1) {$\bx$};
\node[text = virid1] (gx) at (-5,1) {$\gx$};
\node[text = br_green] (bu) at (5,1) {$\bu$};
\node[text = virid1] (gu) at (-4,1) {$\gu$};
\node[text = br_blue] (fx) at (-6,2) {$\pO_{\X}$};
\node[text = virid_purp] (fu) at (6,2) {$\pO_{\U}$};
\node[text = virid1] (fa_xu) at (-2,2) {$\pO_{\A|\X,\U}$};
\node[text = br_green] (fy_axu) at (2,2) {$\pO_{\Y|\A,\X,\U}$};
\node (fa) at (-1,4) {$\pO_{\A}$};
\node (fy) at (1,4) {$\pO_{\Y}$};
\node (siga) at (-3,5) {$\siga$};
\node (sigy) at (3,5) {$\sigma_y^2$};
\node[text = virid1] (aa) at (-3.5,0) {$\alpha_a$};
\node[text = br_green] (ay) at (3.5,0) {$\alpha_y$};
\node (bamp) at (0,8) {$(\bhatx(\X),\bhatx(\emptyset))$};
\node[text = br_green] (Ba) at (6,1) {$\ba$};
\node (n) at (0,0) {$n$};
\draw[->] (mu_x) to [out = 70, in = -70] (fx);
\draw[->] (mu_u) to [out = 110, in = -130] (fu);
\draw[->] (sig_x) to [out = 110, in = -130] (fx);
\draw[->] (sig_u) to [out = 70, in = -50] (fu);
\draw[->] (gx) to [out = 70, in = -180] (fa_xu);
\draw[->] (bu) to [out = 110, in = 0] (fy_axu);
\draw[->,darj_red] (gu) to [out = 0, in = -100] (fa_xu);
\draw[->] (bx) to [out = -180, in = -90] (fy_axu);
\draw[->] (sig_ea) to [out = 150, in = -90] (fa_xu);
\draw[->] (sig_ey) to [out = 30, in = -90] (fy_axu);
\draw[->,darj_red] (fa_xu) to [out = 110, in = -150] (fa);
\draw[->,darj_red] (fa_xu) to [out = 0, in = -90] (fy);
\draw[->] (fy_axu) to [out = 70, in = -30] (fy);
\draw[->] (fx) to [out = 110, in = -180] (fa);
\draw[->] (fx) to [out = 0, in = 130] (fy);
\draw[->] (fu) to [out = 70, in = 0] (fy);
\draw[->] (fu) to [out = -180, in = 50] (fa);
\draw[->,darj_red] (fa) to [out = 110, in = 0] (siga);
\draw[->,darj_red] (fy) to [out = 70, in = -180] (sigy);
\draw[->,darj_red] (siga) to [out = 10, in = -100] (bamp);
\draw[->] (gx) to [out = 120, in = - 160] (bamp);
\draw[->,darj_red] (gu) to [out = 120, in = - 150] (bamp);
\draw[->] (bu) to [out = 60, in = - 20] (bamp);
\draw[->] (bx) to [out = 60, in = - 20] (bamp);
\draw[->] (sig_x) to [out = 110, in = -160] (bamp);
\draw[->] (sig_u) to [out = 70, in = -20] (bamp);
\draw[->] (aa) to [out = 20, in = -90] (fa_xu);
\draw[->] (ay) to [out = 160, in = -90] (fy_axu);
\draw[->] (Ba) to [out = 110, in = 0] (fy_axu);
\draw[->] (n) to [out = 90, in = -90] (bamp);
\draw[->,darj_red] (sigy) to [out = -190, in = -90] (bamp);
\end{tikzpicture}
}
\caption[DAG describing bias amplification simulation]{A DAG representing the simulation of changing $\gu$. The paths from $\gu$ to the outcome [$(\bhatx(\X),\naive)$] are shown with red lines. The colour coding groups the parameters with the (conditional) distributions which we directly generate using the structural equations. Blue is used for those functionals and parameters associated with $\X$, purple is used for those associated with $\U$, yellow with the treatment distribution $\A|\X,\U$, and green for the conditional outcome distribution $\Y|\A,\X,\U$.}\label{sim dag 2}
\end{figure}

Suppose the outcome variable is bias amplifcation as described in equation \eqref{eq bamp 1} which is a functional of the jointly estimated regression coefficients from the conditional model and naive model, $(\bhatx(\X),\bhatx(\emptyset))$. To represent the asymptotic distributions of the target estimators we can leverage their asymptotic means and variances. The means can be found by rearranging equations \eqref{bhatx plim} and \eqref{naive plim}, whereas the asymptotic variance can be found by leveraging theory regarding misspecified regression functionals in \cite{buja2019models1} and results from \cite{stokes2022causal}. Since the structural equations are linear and all the simulated variables are normal, this implies that a linear regression is well-specified (in the sense of \cite{buja2019models1,buja2019models2}) for the conditional outcome distributions $\Y|\A,\X$ and $\Y|\A$. This implies the following asymptotic variances for $\sqrt{n}(\bhatx(\X) - \ba)$ and $\sqrt{n}(\bhatx(\emptyset) - \ba)$ respectively:
\begin{align*}
Var(\sqrt{n}(\bhatx(\X) - \ba)) &=   \frac{E[Var(\Y|\A,\X)\upsilon_{\A|\X}^2]}{(\sigma_a^2 - \gx^2\sigx)^2}, \text{and}\\
Var(\sqrt{n}(\bhatx(\emptyset) - \ba)) &= \frac{E[Var(\Y|\A)\upsilon_{\A|\emptyset}^2]}{(\sigma_a^2)^2},
\end{align*}
where $\upsilon_{\A|\X}^2$ and $\upsilon_{\A|\emptyset}^2$ are the population squared residuals from the regression of the treatment on $\X$ plus a constant and the regression of the treatment on a constant respectively. Using standard multivariate normal theory the conditional variances, $Var(\Y|\A,\X)$ and $Var(\Y|\A)$, can be expressed in terms of matrix operations on the variance-covariance matrix shown below:
\begin{align}
    \boldsymbol{\Sigma}_{\Y,\A,\X} &= \begin{vmatrix}
    \sigma_y^2 &\sigma_{y,a} & \sigma_{y,x}\\
    \sigma_{y,a} & \siga & \sigma_{a,x}\\
    \sigma_{y,x} & \sigma_{a,x} & \sigx
    \end{vmatrix},\label{sig yax}
\end{align}
where $\sigma_{y,a} = \ba\siga + \onep(\bx\odot\gx\odot\sigx) + \bu\gu\sigu$,$\sigma_{y,x} = \ba\onep(\gx\odot\sigx) + \onep(\bx\odot\sigx)$, and $\sigma_{a,x} = \onep(\gx\odot\sigx)$. For simplicity, we assume the regressors $\X$ are mutually independent but this example can be extended to the dependent case. Thus the asymptotic variances can be expressed in terms of the manipulable parameters (including the sample size $n$) and the marginal treatment and outcome variances $\siga$ and $\sigy$. Together the asymptotic means and variances allow us to represent a DAG for the asymptotic distribution of the joint distribution of our outcome, $(\bhatx(\X),\naive)$ as seen in Figure \ref{sim dag 2}.

The DAG in Figure \ref{sim dag 2} is unusual in that it contains parameters and arrows from which will be deterministic unless the simulation also specifies some random distribution over the manipulable parameters. Here we build on Dawid's work on influence diagrams \cite{dawid2002influence} which provides a skeleton for a coherent graphical representation of such structures in the context of causal relationships. Although not explored in this text, the functional nodes used in the influence diagram framework may be very useful for representing simulation experiments like the previous Elements of Statistical Learning \cite{hastie2009elements} example.\\

When examining the simulation experiment DAG in Figure \ref{sim dag 2}, we can see that intervention on $\gu$ causes changes along three pathways to the outcome. The first path is directly from $\gu$ to the outcome via the asymptotic mean $(\gu \rightarrow (\bhatx(\X),\naive))$. The second is indirectly through the conditional treatment distribution which changes the marginal variance of the treatment $(\gu \rightarrow \pO_{\A|\X,\U} \rightarrow \pO_{\A} \rightarrow \sigma_a^2 \rightarrow (\bhatx(\X),\naive)$. The third is also indirectly through the marginal outcome variance ($\gu \rightarrow \pO_{\Y} \rightarrow \sigy \rightarrow (\bhatx(\X),\naive))$. This gives us a new perspective on the two bias amplification simulations performed in \cite{stokes2022causal}. When we change $\gu$ to $\gu^\prime$ holding all other simulation parameters constant, we estimate a total effect of changing $\gu$ which includes the effects through both direct and indirect paths. This indirect effect through the marginal treatment variance in particular weakens the bias amplifying effect of the control variables $\X$ by increasing the marginal variance of the treatment, $\siga$. Our proposed alternative simulation more closely addresses the question of bias amplification because it held the marginal treatment and outcome variance constant, in addition to all other simulation parameters, by using the error variance to absorb the changes. As argued in \citep{stokes2022causal} this approach may be especially important in simulation studies built on real data sets since quantities like the marginal variance are known and fixed in real data sets. Finally, from the perspective of causal inference, this second simulation experiment corresponds to the estimation of a natural direct effect. \\

A natural direct effect estimate requires that the indirect pathways remain constant and that the mediators must be unconfounded \cite{vanderweele2013three}. There are many ways that this can be done in this DAG. A natural way to block the indirect path would be by blocking through $\sigma_a^2$ and $\sigma_y^2$. It must be noted that $\pO_{\A}$, $\pO_{\A|\X,\U}$, and $\pO_{\Y}$ have also been changed. To ensure no information leaks, one would need to block several other backdoor paths. One sufficient set of blocking variables is: $(\gx,\sigx,\sigu,\ba,\bu,\bx,\siga,\sigma_y^2)$. There are many other combinations of nodes which can block the open paths in this DAG, but this particular set of variables maximizes the number of blocking variables which are directly manipulable parameters in our parameterization for the data generating process. By choosing the blocking set which maximizes the number of directly manipulable parameters we reduce much of the blocking to simply holding simulation parameters constant across treatments.\\

When we run simulations conditional on some set of parameters held constant, we are ultimately estimating simulation causal effects conditional on the said set of parameters. In such cases, one must be careful to ensure that we are not changing any parameters which directly impact the outcome, even if they do not lie on any confounding pathways with respect to the effect we are interested in estimating unless we would like to consider them part of the exposure. In this case, we can see that the sample size, $n$, directly impacts the outcome and thus should be held constant. This gives us the following vector of quantities needed to hold constant - $(\gx,\sigx,\sigu,\ba,\bu,\bx,\siga,\sigma_y^2,n)$ -  which we can decompose into a group of directly manipulable variables $(\gx,\sigx,\sigu,\ba,\bu,\bx,n)$ and downstream quantities $(\siga,\sigma_y^2)$. To fix the directly manipulable variables we simply need to leave their values fixed and unchanged when we change the value of $\gu$. To hold the downstream variables constant we need to find variables which contribute to the marginal variance, which are not required to remain fixed. In this case, we have the error terms $\sigma_{\epsilon_a}$ and $\sigma_{\epsilon_y}$, which impact the marginal variances through the paths $(\sigma_{\epsilon_y} \rightarrow \pO_{\A|\X,\U} \rightarrow \pO_{\A} \rightarrow \siga)$ and $(\sigma_{\epsilon_y} \rightarrow \pO_{\Y|\A,\X,\U} \rightarrow \pO_{\Y} \rightarrow \sigma^2_y)$. We use the error variances to absorb changes in the marginal variances and hold them constant. We can see through the DAG that the error variances do not impact the outcome in any other way except through the marginal variances and thus can vary them freely.\\

Now consider a comparison of the total and direct effect simulation experiments for changing $\gu$ to $\gu^\prime$. In the following Figure \ref{fig:bhatx gu gu2} we will compare the results of the two different simulation approaches comparing bias amplification when $\gamma_u = 0.3$ to when $\gamma_u = 0.55$. The other parameters are described in Tables \ref{table ch5 gu1} and \ref{table ch5 gu 2}.

\begin{table}[H]
\centering
\caption{Simulation Parameters 1}
\begin{tabular}{@{}cccc@{}}
\toprule
                                                          & \multicolumn{3}{c}{\textbf{Variable}}                 \\
\multicolumn{1}{c|}{\textbf{Structural Equations}}        & $\A$                 & $\X$ & $\U$ \\ \midrule
\multicolumn{1}{c|}{$\Y$ ($\boldsymbol{\beta}_{\cdot}$)}  & .2                   & -0.05     & .3\\
\multicolumn{1}{c|}{$\A$ ($\boldsymbol{\gamma}_{\cdot}$)} & $\cdot$ & 0.6    & (0.3,0.55)\\ \bottomrule
\end{tabular}\label{table ch5 gu1}
\end{table}

\begin{table}[H]
\centering
\caption{Simulation Parameters 2}
\begin{tabular}{@{}c|cccc@{}}
\toprule
                  & \multicolumn{4}{c}{\textbf{Variable}}                                                            \\
\textbf{Quantity} & $\Y$    &  $\A$      & $\X$      & $\U$           \\ \midrule
Intercept         & 0       & 0     &  $\cdot$           & $\cdot$\\
Variance          & (1, 1.04)       & (1, 1.21)          & 1      & 1\\ \bottomrule
\end{tabular}\label{table ch5 gu 2}
\end{table}

\begin{figure}[H]
    \centering
    \includegraphics[scale = 0.25]{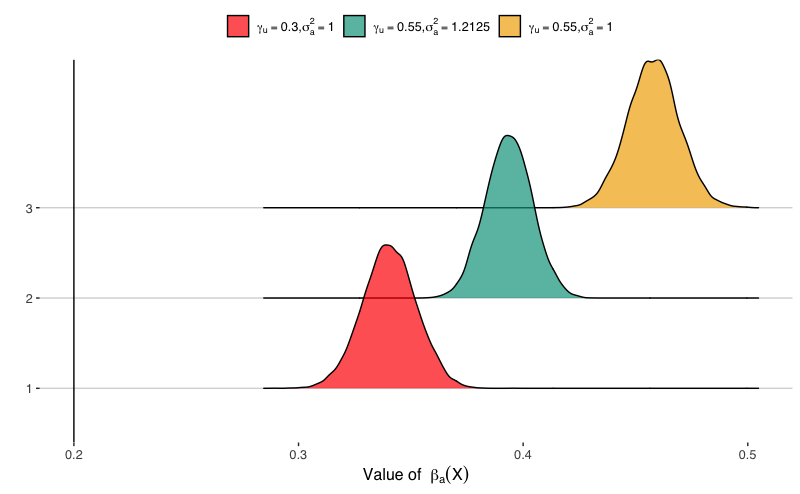}
    \caption[Effect of changing $\gu$ and allowing marginal variances to float vs. held fixed]{Simulation Results for $\bhatx(\X)$. In red is the control arm, where $\gamma_u = 0.3$. In teal is the total effect of $\gamma_u$ experiment treatment arm where $\gamma_u = 0.55$. In this simulation experiment the marginal variance of the treatment and the outcome increase to 1.21 and 1.04 respectively from 1. In yellow is the direct effect of $\gamma_u$ experiment where the marginal variance of the treatment and outcome are held constant.(n =10,000 with 10,000 replications)}
    \label{fig:bhatx gu gu2}
\end{figure}
Above in Figure \ref{fig:bhatx gu gu2}, we can see simulation results for the conditional estimator $\bhatx(\X)$ from the two approaches. On the bottom row in red is the control arm of the simulation experiment where $\gamma_u = 0.3$. We can see that this estimator is biased and well above the true parameter $\beta_a = 0.2$ (shown with the vertical black line). The probability limits from equation \eqref{bhatx plim} for the control estimator is $\beta_a + \bu\gu\sigu/(\siga - \gx^T\sigx\gx) = 0.341$ which is identical to the empirical average to three decimal places. In teal we see the simulation described by the graph in Figure \ref{fig:ext dag 2}. This shows the total effect of increasing $\gamma_u$. In yellow we see the direct effect simulation, where we control the indirect paths by fixing the marginal outcome and treatment variances.\\

\begin{figure}[H]
    \centering
    \includegraphics[scale = 0.25]{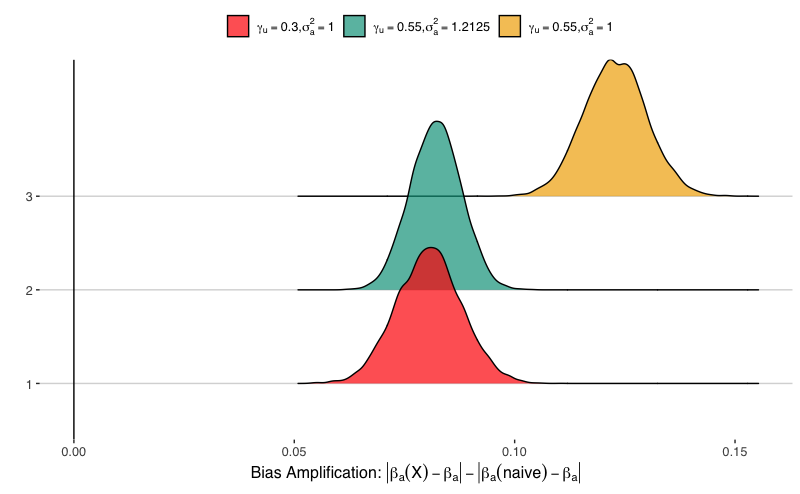}
    \caption[Bias Amplification simulation comparing fixed and floating marginal variance interventions]{This graphs plots the same simulation results as Figure \ref{fig:bhatx gu gu2}. The x-axis is now the additional absolute bias of the conditional estimator compared to the naive.}
    \label{fig: gu bamp}
\end{figure}

The two versions of the experiment are not equivalent. The total effect of changing $\gamma_u$ from 0.3 to 0.55 increases the average value of $\bhatx(\X)$ from 0.341 to 0.391 (in teal in Figure \ref{fig:bhatx gu gu2}). The direct effect of changing $\gamma_u$ from 0.3 to 0.55, or the direct effect of intervening on $\U \rightarrow \A$, increases the average value of $\bhatx(\X)$ from 0.341 to 0.458. This is quite a large difference. When we look at the effect on the additional bias of $\bhatx(\X)$ compared to $\naive$ the difference between the two simulation experiments is even more stark (Figure \ref{fig: gu bamp}).
In Figure \ref{fig: gu bamp} we see the additional absolute bias of the conditional estimator compared to the naive estimator ($|\bhatx(\X) - \beta_a| - |\naive - \beta_a|$) for the control arm and the two versions of the simulation intervention. Comparing the results in teal, where we increase $\gamma_u$ but allow the marginal variances to change, we see that the additional bias of the conditional estimator hardly increases (0.082 compared to 0.081). However, when we hold the marginal variance of the treatment constant, increasing $\gamma_u$ increases the bias amplification significantly from 0.081 to 0.123. As we argued in \cite{stokes2022causal}, this may explain why simulation results from \cite{meyers2011} downplayed the risks of bias amplification and did not easily accord with the theory and probability limits developed at the time, notably \cite{pearl2012class,pearl2011invited}. The indirect effects of intervening on $\gu$ reduce the overall bias and bias amplification by reducing the capacity of $\X$ to be bias amplifiers, which explains the discrepancy in the two simulations. In general, neither simulation experiment is incorrect, but just as total and direct effects answer different causal questions these simulation experiments answer different questions. Problems arise when we incorrectly design experiments for the research question at hand.\\

\begin{figure}[H]
    \centering
    \includegraphics[scale = 0.25]{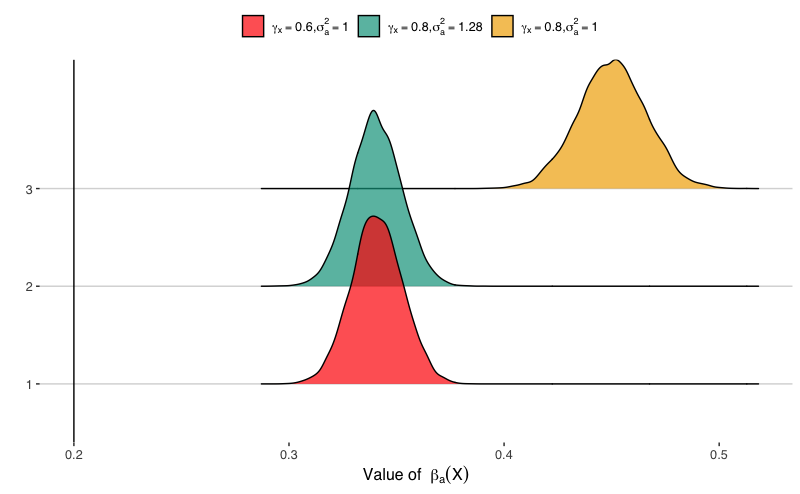}
    \caption[The impact of intervening on $\A \rightarrow \X$ on $\bhatx$ (fixed vs. floating]{The experiment of changing the relationship of $\A \rightarrow \X$ by changing $\gx$. The red shows the control, the teal the exposure arm where we allow the treatment variance to vary, and the yellow is the exposure where we use the independent error of the treatment to hold the treatment marginal variance constant.}
    \label{fig:gx bx}
\end{figure}

Consider another simulation experiment. The control arm distributions remains the same, and this time we will increase the effect of $\X \rightarrow \A$ through the parameter $\gx$. A DAG representing this experiment is identical to the one in Figure \ref{sim dag 2} after swapping the places of $\gx$ and $\gu$. That is $\gx$ has one direct effect on the outcome and two indirect paths which go through the marginal outcome and treatment variances. In the control arm of this simulation experiment, we set $\gx$ to 0.6 and then increase it to 0.8 and run both versions of that experiment. From the theory developed in \cite{stokes2022causal}, we know that bias amplification is very related to the hyperbolic effect of the variance explained by the control variables $\X$. The smaller that $\siga - \gx^T\sigx\gx$ is in the control arm, the larger differences we will expect between the total and direct effect version of the experiment given the nature of a hyperbolic curve (Figure \ref{fig:gx bx}).
In Figure \ref{fig: gu bamp} we see that when we increase $\gx$ but allow the marginal variance to increase, $\bhatx(\X)$ remains unchanged compared to the control. In fact, the probability limits from  equation \eqref{bhatx plim}
are identical for the two estimators, the control (in red) and the total effect of $\gx$ treatment arm (in teal). We can see this below where the probability limit in the control arm is:
\begin{align}
    \plimn \bhatx(\X;\gx) = \beta_a + \frac{\bu\gu\sigu}{\siga - \gx^T\sigx\gx},
\end{align}
and in the treatment arm we have:
\begin{align}
    \plimn &\bhatx(\X;\gx^\prime)= \beta_a + \frac{\bu\gu\sigu}{\sigaprime - \gx^{\prime T}\sigx\gx^{\prime}}\label{eq rm same 1}\\
    &= \beta_a + \frac{\bu\gu\sigu}{\siga + \onep({\gx^\prime}^2 - \gx^2)\odot\sigx - {\gx^\prime}^t\sigx\gx^\prime}\label{eq rm same 2}\\
    &= \bhatx(\X;\gx),\label{eq rm same 3}
\end{align}
since since $\sigaprime = \siga + \onep({\gx^\prime}^2 - \gx^2)\odot\sigx$ when the $\X$'s are orthogonal. Thus when we look at the bias amplification in this version of the experiment, the only change that we are observing is in the naive estimator, which in general is not the same in the two treatment arms since $\frac{\onep(\bx\odot\gx\odot\sigx)}{\siga} \neq \frac{\onep(\bx\odot\gx^\prime\odot\sigx)}{\sigaprime}$ and $\frac{\bu\gu\sigu}{\siga} \neq \frac{\bu\gu\sigu}{\sigaprime}$. When we hold the marginal variance constant however, changing $\gx$ can have a large effect on $\bhatx(\X)$ as seen within Figure \ref{fig:gx bx} in yellow. Compared to the control arm, the direct effect of increasing $\gx$ from 0.6 to 0.8 increases the average of $\bhatx(\X)$ from 0.341 to 0.450 which is a relatively large impact.
\begin{figure}[H]
    \centering
    \includegraphics[scale = 0.25]{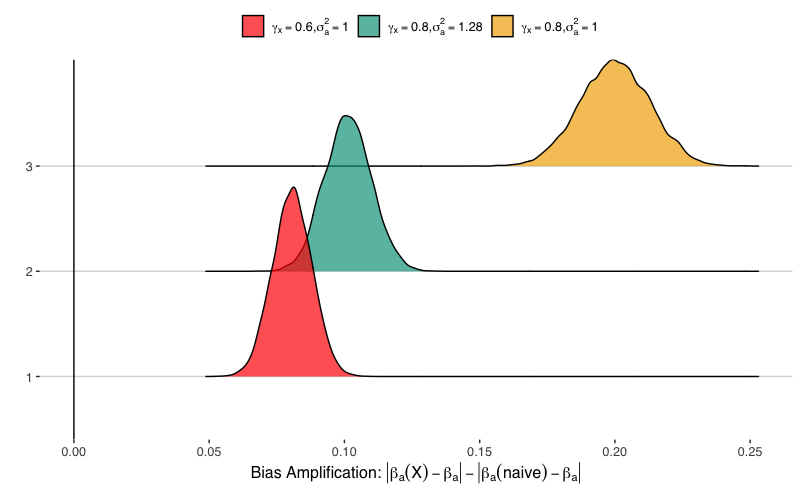}
    \caption[Bias Amplification as $\gx$ changes]{Here we see again the experiment of changing $\gx$, where red is the control arm and teal and gold are the exposures while allowing the marginal variance to float and holding it constant respectively.}
    \label{fig:gx bamp}
\end{figure}
In Figure \ref{fig:gx bamp}, we see the additional absolute bias of the conditional estimator compared to the naive estimator. In the control arm, we have an average bias amplification of 0.081 (in red). When we increase $\gx$ to 0.8 but allow the marginal variances to vary, the average bias amplification increases a little to 0.10. As show in Figure \ref{fig:gx bx}, this is entirely because the naive estimator has become less biased since the conditional estimator remains unchanged. When we hold the marginal variance of the treatment and outcome constant the additional bias in the conditional estimator is much larger, increasing to 0.20.\\

The difference between the two simulation experiments, and whether or not we fix the marginal variance, underscores why early work in bias amplification from the theoretical side, notably \cite{pearl2012class} and \cite{pearl2011invited}, took the threat of bias amplification more seriously than was suggested by the simulation studies conducted by Myers et al \citep{meyers2011}. In his invited commentary to \citep{meyers2011}, for example, Pearl \citep{pearl2011invited} discusses how the bias due to the treatment variance remaining term ($\sigma_a^2 - \gamma_{x_1}^2\sigma_{x_1}^2 -\dots - \gamma_{x_k}^2\sigma_{x_k}^2$) ``increases monotonically" with the number of covariates included in the model. This is only consistent with the implicit assumption that the marginal variance of the treatment remains fixed when adding variables, since otherwise the denominator remains unchanged as shown in equations \eqref{eq rm same 1} through \eqref{eq rm same 3}. The point is not that this analysis or that direct effect analysis is inherently correct, as the focus of this article is simulation and not bias amplification itself. Rather, it should not be surprising that these two different types of causal effect analyses disagree. Often when we investigate formulas we are typically implicitly conducting static experiments where we hold parts of the formula constant while allowing other parts to vary. In some applied contexts this is likely the natural model of analysis. In a sensitivity analysis with real data, for example, the amount of total treatment and outcome variance is implicitly fixed by the data generating process and realized data and does not change when we make different assumptions about the unmeasured confounding structure. As seen in this section, however, simulation experiments do not necessarily conduct this kind of analysis which can lead to different or potentially incorrect conclusions when treated equivalently. As with other parts of causal inference, which effect is of interest is ultimately contextual and dependent on the question. As shown in this article, we can design simulation experiments to match either of these questions and use many of the standard causal inference tools to do so.

\subsection{Holding quantities constant implies constraints}\label{sbsec:ch5 constraints}

In systems of equations, holding some quantities constant while varying others will often imply constraints on the way in which other parameters may be varied. This is especially true of downstream variables, those which are not directly manipulable given the parameterization of the joint density, since they may be a non-trivial function of the directly manipulable. Even directly manipulable parameters may constrain others depending on joint density or constraints relating to other desired properties of the simulation (say that the outcome distribution has a mean in some range $[a,b]$ where $a < b$). When one attempts to estimate a direct effect via simulation, constraints on the parameter space are likely to arise.\\

Consider again the first simulation experiment from the previous section, where we varied $\gu$ subject to holding the marginal variance of the treatment constant. This implies the constraints on $\gu$. In that case, we set the marginal variance of the treatment to 1, $\siga = 1$, as well as the marginal variances of $\X$ and $\U$. Recall that we used the variance of the independent error $\epsilon_a$ to hold the treatment variance constant. Since this variance must be non-negative it implies that:
\begin{align}
    |\gu| \leq (1 - \gx^2)^{\frac{1}{2}}.
\end{align}
Holding the marginal variance of the outcome constant, as we did in the previous section, additionally implies the constraint:
\begin{align}
    \gu \leq \frac{1 - \ba^2 - \bu^2 - \bx^2 - 2\ba(\onep\bx)}{2\ba\bu}.
\end{align}
Notice that these constraints depend on the values of the other parameters. Not only may our conditional simulation results give us different results given different parameter sets, but what interventions we can perform is determined by the conditioning set as well. In this particular example, given the parameters as described in Tables \ref{table ch5 gu1} and \ref{table ch5 gu 2}, this implies $\gu \in (-0.8, 0.8)$.\\

Depending on what we intend to hold constant and the complexity of the structural equations, parameters or sets of parameters may be subject to many constraints. In addition, holding marginal variances constant requires either deriving the variance formulas or performing numerical optimization. This certainly makes running simulations more complicated and time-consuming. However, parameter space constraints can clarify the space of possible experiments. In the unconstrained version of the $\gu$ experiment, the parameter space $(-\infty,\infty)$ is so large that there is no simple and efficient way to explore the total space. In practice this means that many applied researchers will simply choose a small subset of this parameter space which seems reasonable to perform their experiments. The range $\gu \in (-0.8,0.8)$ is relatively easy to discretize in a meaningful way and even a coarse first pass may reveal whether a finer grid of this parameter space is necessary. 

\begin{figure}[H]
    \centering
    \includegraphics[scale = 0.25]{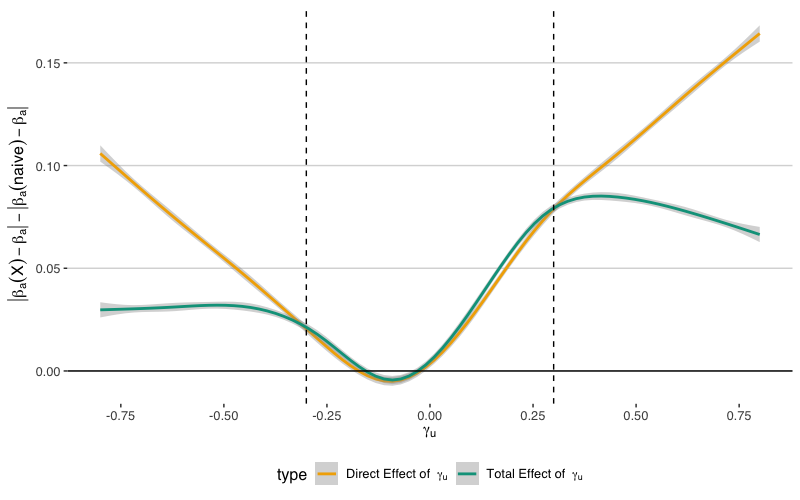}
    \caption[Bias Amplification vs $\gu$ comparing fixed and floating marginal treatment variance experiments]{The x-axis shows the varied parameter $\gu$ and the y-axis is the additional bias of the conditional estimator compared to the naive estimator. The horizontal zero line is the point where there is no bias amplification. We vary $\gu$, this time on a fine grid from -0.8 to 0.8 since these are the constraints in the fixed variance experiment. The dashed vertical lines show the control case of $\gu = 0.3$ and also $\gu =-0.3$. At $\gu \in \{-0.3,0.3\}$ the variance in both versions of the experiment is equal. Otherwise, the marginal variance of the treatment is not equal and as see in the graph the results diverge, particularly outside of the two dashed lines. The yellow line shows the experiment where marginal variance is fixed to 1 throughout and teal shows the floating variance comparison.}
    \label{fig:gu vary}
\end{figure}
In Figure \ref{fig:gu vary}, we discretize the interval $(-0.8,0.8)$ into 10,000 equally spaced points. For each value of $\gamma_u$ we estimate the conditional and naive estimator twice. First, we hold the marginal variances of the treatment and outcome constant (direct effect) and then we allow these quantities to vary freely (total effect). We then plot the value of $\gamma_u$ against the additional bias in the conditional estimator for both versions of the experiment. Notice that at $\gamma_u \in \{-0.3,0.3\}$ (the vertical dashed lines), the direct and total effect agree. This is because $\gamma_u = 0.3$ is the reference value for the experiment, so the variances are equivalent at this level. There is equivalence at $\gamma_u = -0.3$ since $\gamma_u$ enters the variance equation as a squared term, In between the dashed lines, both versions of the experiment nearly agree since for small levels of $\gamma_u$ the impact on the marginal variances is also small. Outside of the dashed lines however, the experiments diverge. On the right side we see more context for the experiment we ran in the previous section increasing $\gamma_u$ from 0.3 to 0.6 (Figure \ref{fig: gu bamp}). We see over this region, that the total effect version of the experiment begins to decrease the amount of bias amplification, whereas the direct effect version continues to increase linearly. This effect where the total and direct effect diverge away from the reference value will depend on the reference value itself since this changes the effective marginal treatment and outcome variances at the different levels of $\gamma_u$ in the total effect experiment.\\

When the structural equations are additive in the error term, constraints can be expressed in terms of the variance-covariance matrix. In this case we have the following constraints:
\begin{align}
  0 \leq  \sigma_{\epsilon_y}^2 &= \sigma_y^2 -  [\ba,\bx,\bu]Var([\A,\X,\U])[\ba,\bx,\bu]^{\boldsymbol{T}}\, and\\
  0 \leq \sigma_{\epsilon_a}^2 &= \siga - [\gx,\gu]Var([\X,\U])[\gx,\gu]^{\boldsymbol{T}},
\end{align}
where both of the variance-covariance matrices are sub-components of:
\begin{align}
    \boldsymbol{\Omega} &= \begin{vmatrix}
\sigma_y^2 &\frac{\rho_{\Y,\A}}{\sigma_y\sigma_a} & \frac{\rho_{\Y,\X}}{\sigma_y\sigma_a} & \frac{\rho_{\Y,\U}}{\sigma_y,\sigma_u}\\
\frac{\rho_{\Y,\A}}{\sigma_y\sigma_a} &\siga & \frac{\rho_{\A,\X}}{\sigma_a\sigma_x} & \frac{\rho_{\A,\U}}{\sigma_a,\sigma_u}\\
\frac{\rho_{\Y,\X}}{\sigma_y\sigma_x}&\frac{\rho_{\A,\X}}{\sigma_a\sigma_x} & \sigx & \frac{\rho_{\X,\U}}{\sigma_x\sigma_u}\\
\frac{\rho_{\Y,\U}}{\sigma_y\sigma_u} & \frac{\rho_{\A,\U}}{\sigma_a\sigma_u} & \frac{\rho_{\X,\U}}{\sigma_x\sigma_u} & \sigu
\end{vmatrix}.
\end{align}
When working with linear models especially, it may be convenient to express the constraints in terms of these correlation matrices and marginal variances, particularly when the structural equations get more complicated and the number of structural equations grows. As discussed in \cite{stokes2022causal}, under less restrictive forms of structural equations, the limit of OLS estimators can be represented in terms of marginal variances and partial correlations. The partial correlations can then in turn be reduced to the marginal pairwise correlation matrix.\\

Working with the underlying pairwise correlation matrix may also make it easier to elicit domain knowledge since structural parameters do not typically have convenient interpretations and crucially depend on the other variables (and their function form) in the structural equation or model. The difference is especially true when working with real data. The pairwise correlations of all observable variables is always estimable independent of any unmeasured variables, whereas structural equations are only estimable under unmeasured equations under restrictive assumptions. Treating $\U$ as unmeasured and $(\Y,\A,\X)$ as measured, the bolded quantities in $\Omega$ can be estimated from the samples of the joint observed distribution, whereas the parameters in red cannot.
\begin{align}
    \boldsymbol{\Omega} &= \begin{vmatrix}
\boldsymbol{\sigma_y^2} &\boldsymbol{\frac{\rho_{Y,A}}{\sigma_y\sigma_a}} & \boldsymbol{\frac{\rho_{Y,X}}{\sigma_y\sigma_a}} & \color{darj_red}{\frac{\rho_{\Y,\U}}{\sigma_y,\sigma_u}}\\
\boldsymbol{\frac{\rho_{Y,A}}{\sigma_y\sigma_a}} &\boldsymbol{\sigma_a^2} & \boldsymbol{\frac{\rho_{A,X}}{\sigma_a\sigma_x}} & \color{darj_red}{\frac{\rho_{\A,\U}}{\sigma_a,\sigma_u}}\\
\boldsymbol{\frac{\rho_{Y,X}}{\sigma_y\sigma_x}}&\boldsymbol{\frac{\rho_{A,X}}{\sigma_a\sigma_x}} & \boldsymbol{\sigma_x^2} & \color{darj_red}{\frac{\rho_{\X,\U}}{\sigma_x\sigma_u}}\\
\color{darj_red}{\frac{\rho_{\Y,\U}}{\sigma_y\sigma_u}} & \color{darj_red}{\frac{\rho_{\A,\U}}{\sigma_a\sigma_u}} & \color{darj_red}{\frac{\rho_{\X,\U}}{\sigma_x\sigma_u}} & \color{darj_red}{\sigu}
\end{vmatrix}\label{const matrix}
\end{align}
As discussed in \cite{stokes2022causal}, in particular cases like when the simulation is a sensitivity analysis for an unmeasured confounder, the induced constraints can be partially determined by the set of valid $(n+1)\times (n+1)$ correlation matrices, given an estimated $n\times n$ correlation matrix. More explicitly, this is equivalent to extending the inner matrix represented by the black parameters to the full matrix including the red parameters in equation \eqref{const matrix}. There are algorithms, such as the work in \cite{budden2008generation}, which can be leveraged to extend $n\times n$ correlation matrices to valid $(n+1)\times (n+1)$ correlation matrices which may be preferable to solving systems of equations in some cases. Further, there are methods to simulate noise around structured correlation matrices \citep{hardin2013method} which may be used to incorporate the uncertainty due to part of the matrix being estimated. This method can be extended iteratively to incorporate unmeasured confounders of dimension greater than 1.\\

In some cases it may be useful to choose an ordering over constraints that one will apply. In such cases it may be useful to encode this ordering into a graphical model. An extreme example of this can be found in Figure \ref{fig: esl graph}, where the signal fully determines the noise which can be seen as a hard constraint. Choices over different ordering of constraints may lead to different experiments and experimental designs.

\subsubsection{Constraints in the absence of theory}
In the previous section, we discussed how constraints arise when estimating direct effects via simulation and provided an example in the context of linear models with linear, normal structural equations. In such a case, it is not difficult to derive a closed form expression for the marginal variances in terms of the error terms, which we rearranged to find the parameter constraints. In some cases, the functional we want to hold constant may have an unknown or partially unknown relationship to the directly manipulable parameters. In such cases, we may still be able to perform the desired simulations.\\

\sloppy Let $\theta_{m}$ be the set of directly manipulable parameters, where in our canonical bias amplification example $\theta_m = (\alpha_y,\beta_a,\beta_u,\bx,\alpha_a,\gu,\gx,\mu_u,\boldsymbol{\mu_x},\sigu,\sigu,\sigx,\sigma_{\epsilon_y}^2,\sigma_{\epsilon_y}^2, n)$. Let $\boldsymbol{\psi}(\pO;\theta_m)$ be the functional or set of functionals we would like to hold constant, which we will suppose are the marginal variances $[\sigma_a^2(\pO;\theta_m),\sigma_y^2(\pO;\theta_m)]$ for illustrative purposes. In some cases one might be able to use the simulation DAG and/or the structural equations to reduce the set of variables to a subset, i.e we can represent $\boldsymbol{\psi}(\pO;\theta_m)$ as $\psi(\pO;\theta_s)$ where $\theta_s \subset \theta_m$. This subset may then be partitioned into two groups - parameters which need to be held constant for the desired simulation ($\theta_s^c$) and those which may be varied freely ($\theta_s^f$). We can then write the desired constant functional as $\boldsymbol{\psi}(\pO;\theta_s^f,\theta_s^c)$. The problem of holding this function constant can then be recast as an optimization problem in $\theta_s^f$.
\begin{align*}
\theta_s^{f,opt} = \underset{\theta_s^f \in \Theta_s^f}{\arg\min} E_{\pO}[\mathcal{L}(\psi^0,\hat{\psi}(\pO;\theta_s^f,\theta_s^c)]
\end{align*}
where $\psi^0$ are the desired constants and $\mathcal{L}(\cdot,\cdot)$ is a loss function. When $\theta_s^f$ is sufficiently well-behaved many strategies are available to solve the optimization problem including the usual gradient-based solvers, for example, this optimization problem is orders of magnitude easier than those in the real-world since we have access to the data generating mechanism. This means we can always generate new independent sets of data to increase the precision of the optimization, to test hypotheses on restrictions to $\theta_s^f$, or even to learn features of the unknown function itself.\\

In our example, we will show that without using full knowledge of the expressions which determine the marginal variances, we can reduce the problem of holding these variables constant to a low-dimensional optimization problem. First, we can remove $n$ from $\theta_m$ since it has no paths to either $\sigma_a$ or $\sigma_y$. The marginal treatment variance also precludes all of the manipulable parameters which are parents of $\pO_{\Y|\A,\X,\U}$ in the DAG in Figure \ref{sim dag 2} since the treatment is generated prior to the conditional outcome in the structural equations. Statistical knowledge may allow us to reduce these subsets further. Knowing that all the variables are generated as normal variables and their dependencies allows us to reduce this set further to not include any of the mean or intercept terms since they will not change the variances. This leaves us with the following functionals and sets:
\begin{align*}
\sigma_a^2(\pO;\theta_s^f,\theta_s^c) &= f_1(\sigma_{\epsilon_a};\gx,\gu,\sigu,\sigx)\\
\sigma_y^2(\pO;\theta_s^f,\theta_s^c) &= f_2(\sigma_{\epsilon_a},\sigma_{\epsilon_y};\gx,\gu,\sigu,\sigx,\ba,\bu,\bx,\siga)
\end{align*}
In this case, we see that $\sigma_a^2$ is only a functional of one parameter which does not belong to the set of variables held constant, $\sigma_{\epsilon_a}$. The marginal outcome variance, $\sigma_{\epsilon_y}$ is a functional of two variables not held constant, supposing we do it sequentially after holding $\sigma_a^2$ constant. Using the information in the outcome structural equation we can see that $\sigma_{\epsilon_a}$ only impacts the marginal outcome variance through $\sigma_a^2$. Thus in this problem we can reduce learning the unknown function to a two-step sequential optimization with one unknown. With an appropriate loss function, even if we do not know the true function for the marginal variance in the parameters we can find values for $\sigma_{\epsilon_a}$ and $\sigma_{\epsilon_y}$ which hold the marginal variances constant up to some error. In this case the error will be on the order of a pre-specified tolerance since the underlying unknown functions are well-behaved functions in $\sigma_{\epsilon_a}$ and $\sigma_{\epsilon_y}$. The additional assumption needed is that the optimization problem is well-posed and in some cases further assumptions like uniqueness of the solution may be necessary.\\

As with traditional applications of causal inference, having access to more developed theory increases the number tools at one's disposal for taking a causal approach to simulation experiments. However, we can still use causal inference tools and frameworks to better understand simulation experiments even when we do not have established closed-form results like those which were available in the bias amplification example. In the bias amplification case, however, it is unnecessary to have a closed form expression for the marginal variances in order to hold them constant since with a lower order of knowledge we could reduce the problem to an optimization problem. Graphical models can help us to represent and exploit such knowledge as is the case in other domains where we seek to leverage information from non-statistical domain experts.\\

\section{Discussion}

Simulation experiments are among the most important tools in generating statistical knowledge and understanding of estimators. However, to the author's knowledge, there is no standard analytical framework or agreed upon set of tools to help contextualize or design simulation experiments in practice. We've shown that many ideas and tools from causal inference offer a promising structure and methodological framework in which to ground simulation results. Some of the more challenging questions when evaluating simulation studies are the extent to which they generalize to other contexts. In the causal framework, simulation experiments are estimators for a particular estimand. As discussed in this manuscript, many simulation experiments are estimators for particular conditional expectations, where the conditioning is related to the manipulable parameters held fixed. In one sense, this echoes some of the critiques of simulation experiments, that vis-a-vis theoretical results simulation are less general and can be highly specific to the context at hand. However, the broader causal framework gives us tools and language for putting such conditional estimands in context and literatures like transportability and generalizability might offer routes for more formally repurposing or recontextualizing simulation results to new, but related contexts.\\

Finally, we showed that formal causal tools like graphical models can help ensure that we are designing experiments which are valid estimators of our desired estimands. The two examples we treated in this manuscript - comparing the performance of neural networks under different mean functions and bias amplification both shared the element of model misspecification in common. Model misspecification, as discussed in \cite{buja2019models1,buja2019models2}, can cause estimated parameters and derived functionals of interest to be complicated functionals of the underlying data generating distribution. In particular, the parameters of interest are not ancillary to the covariate distribution. This fact was at the heart of the complication of the two examples treated in this text. When models in simulation experiments are well-specified, as is often the case in many contexts of interest, the complexity reduces greatly and in many cases there may not be a distinction between direct and total causal effects.\\

In both of the examples presented in the text we used statistical theory to help develop graphical and potential outcomes representations of the simulation experiments. In some cases, as in real data causal inference problems, there may be less theory available to describe the relations between parameters and functionals in a statistical experiment. Manipulable parameters will still be determinable directly from the simulation parameters. Further, when constructing a graph there may be reasonable Markov blanket like assumptions which may simplify the graph structure. For example, we might suppose that the set of parameters only impact the outcome functional through a single component of the joint distribution - i.e one assumes a parameter which determines the marginal mean of $\X$ only impacts a downstream function through $\pO_{\X}$. There may be cases where this is insufficient for developing a graphical model rich enough to provide analytical insight into the simulation experiment. In such complicated cases, applications of causal discovery algorithms, algorithms which seek to estimate the underlying causal structure or DAG \citep{glymour2019review}, may be appropriate. The advantage of applying such algorithms and techniques in the setting of simulated data compared to traditional use cases is that the experimenter can always generate more data and even perform manipulations of the underlying structure to increase the amount of information available to feed into the causal discovery algorithm. Testing for both the existence and the direction of an edge in this setting is much simpler setting and, in fact, it may be possible to design new causal discovery algorithms which explicitly take advantage of the added structure inherent in determining the causal graph of a simulated system. The authors additionally hypothesize that there may be techniques from the experimental design and computer experiments literature, such as factorial and space-filling designs for example, which may be used in conjunction with causal discovery techniques to increase the efficiency and accuracy with which one learns the underlying causal structure of a simulation experiment.

\appendix

\section{Additional Figures from E.S.L example}\label{app:esl fig}
This is an additional figure referenced in Section \ref{esl sec} showing a modified version of the simulation experiment in Elements of Statistical Learning.
\begin{figure}[H]
    \centering
    \includegraphics[scale = 0.3]{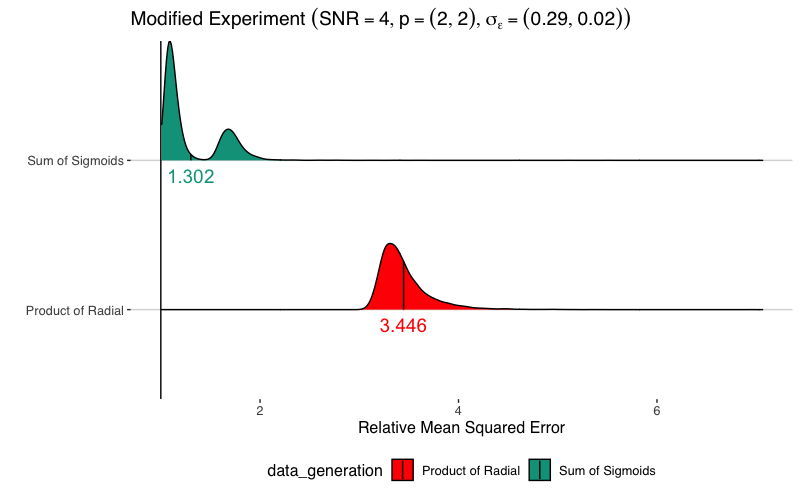}
    \caption{The original experiment but where the radial mean function is only a function of 2 covariates.}
    \label{fig: p2 =2}
\end{figure}

This is a re-creation of the original experiment where $p_2 =2$. This requires that $\sigma_{\epsilon_{radial}} = 0.022$. The results here are much closer to the presented results in Elements of Statistical Learning for 2 nodes. This suggests that likely the experiment was not performed at $p_2=10$ or there is some other discrepancy with the listed protocol.

\begin{figure}[H]
    \centering
    \includegraphics[scale = 0.3]{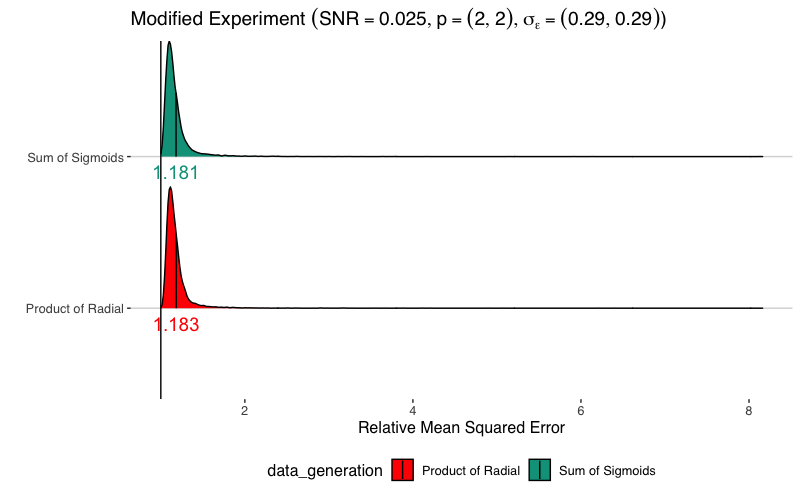}
    \caption[Modified ESL experiment with lower SNR]{Simulation experiment where again the signal-to-noise ratio is held constant by holding both the signal and noise constant. This simulation has much lower signal-to-noise ratio, 0.025, than the previous examples.}
    \label{fig:snr 025}
\end{figure}

Interestingly, in Figure \ref{fig:snr 025} the neural network performs equally well in both treatment arms when the signal to noise ratio is small. This reinforces the notion that the results from these simulation results are inherently conditional. Even when we hold the factors constant that we are uninterested in, in this case the variance of the conditional mean and the signal-to-noise ratio, the simulation results are conditional on those parameters we set and not guaranteed to generalize across the entire distribution of possible parameter choices. Had we only run this experiment we might incorrectly conclude that there is no effect when changing these mean-functions. Consider the simulation experiment in Figure \ref{fig:mse by sigma}, which shows the relative mean-squared error as the irreducible error shifts. All other parameters are taken to be identical to those used to create Figure \ref{fig:snr 025} and in particular the variance of the conditional mean. Therefore the signal-to-noise ratio, remains equal in the two treatments at each level of $\sigma_{\epsilon}$.

\end{document}